\documentclass[aps,prl,10pt,twocolumn,amsmath,amssymb,superscriptaddress,floatfix,nobibnotes]{revtex4-2}

\usepackage{graphicx}
\usepackage{dcolumn}
\usepackage{bm}
\usepackage{epsfig}
\usepackage{color}
\usepackage{stmaryrd}
\usepackage{mathrsfs}
\usepackage[cal=zapfc]{mathalfa}
\usepackage{amsmath}
\usepackage{mdframed}
\usepackage{tcolorbox}
\usepackage[normalem]{ulem}
\usepackage{xcolor}
\usepackage[hidelinks]{hyperref}

\newcommand{\aop}[1]{\hat{a}_{#1}}
\newcommand{\adag}[1]{\hat{a}^{\dagger}_{#1}}

\newcommand{\bra}[1]{\langle{#1}|}
\newcommand{\ket}[1]{|{#1}\rangle}

\newcommand{\ssection}[1]{\phantomsection{\noi \it #1:}}

\newcommand{\expec}[1]{\langle #1 \rangle}

\newcommand{\figuresize}{\columnwidth}

\def\noi{\noindent}

\newcommand{\rref}[1]{Ref.~\cite{#1}}
\newcommand{\fref}[1]{Fig.~\ref{#1}}
\newcommand{\frefp}[2]{Fig.~\ref{#1}~(#2)}

\newcommand{\eref}[1]{Eq.~(\ref{#1})}

\newcommand{\sref}[1]{Sec.~\ref{#1}}

\newcommand{\cref}[1]{chapter~\ref{#1}}

\newcommand{\Cref}[1]{Chapter~\ref{#1}}

\newcommand{\sz}{\hat{S}^{z}}
\newcommand{\sx}{\hat{S}^{x}}
\newcommand{\sy}{\hat{S}^{y}}

\newcommand{\lam}{\lambda}

\newcommand{\hb}{h_{B}}

\newcommand{\ktp}{k_{AB}}

\newcommand{\gbar}{\bar{g}}
\newcommand{\dg}{\Delta g}



\begin{document}

\title{Transition Path Sampling for Rare Events in Non-Markovian Open Quantum Systems}

\author{Ritesh Pant}
\affiliation{Department of Chemistry, University of Texas at Austin, Austin, TX 78712, USA}

\author{Kai Müller}
\affiliation{Institut für Theoretische Physik, Technische Universität Dresden, D-01062 Dresden, Germany}
\affiliation{Clarendon Laboratory, University of Oxford, Parks Road, Oxford OX1 3PU, United Kingdom}

\author{Tarun Gera}
\affiliation{Department of Chemistry, University of Texas at Austin, Austin, TX 78712, USA}

\author{Doran I. G. B. Raccah}
\email{Corresponding author: doran.raccah@utexas.edu}
\affiliation{Department of Chemistry, University of Texas at Austin, Austin, TX 78712, USA}
\begin{abstract}
Rare transitions govern the dynamics of many open quantum systems, but their probabilities and rates are difficult to calculate. Transition path sampling (TPS) of quantum systems has so far been limited to Markovian, perturbative descriptions. Here, we formulate TPS for non-Markovian quantum state diffusion, which is exact for Gaussian baths, extending it to non-Markovian and strongly coupled open quantum systems. For normal-to-superradiant tunneling in the unbalanced open Dicke model, we determine probabilities down to $10^{-8}$ and rates beyond the reach of brute force.
\end{abstract}
\maketitle

\ssection{Introduction}
Rare events are dynamical processes that occur with small probability and yet control the kinetics of many open quantum systems. Examples include transitions between metastable states in driven-dissipative systems~\cite{XiangLiBaiMa2026Switching,MullerStrunz2025Dicke,CardeEtAl2026Switching,GelhausenBuchhold2018DickeBistability,NowoczynMatheySeibold2026LimitCycles,LeeEtAl2025Instanton}, magnetization reversal in molecular magnets~\cite{GaraninChudnovsky1997Tunneling,LunghiEtAl2017AnharmonicPhonons,ThomasCaneschiBarbara1999Mn12}, and hopping of a quantum Brownian particle over the barrier of a double-well potential~\cite{ChristieBolhuisLimmer2025QuantumSSE,ChristieBaekBangJoo2024QuantumSimRates}. Such transitions can also cause bit-flip errors when metastable states are used to encode qubits for quantum information processing~\cite{RegladeEtAl2024CatQubit,SuEtAl2025DoubleWell,LescanneEtAl2020BitFlips}. Free-energy estimates obtained from work measurements in quantum thermodynamics are sensitive to rare events~\cite{JaramilloDengGong2017Work,DengJaramilloHanggiGong2017DeformedJE,LiYaoQiu2026ReverseAnnealing}.

The dynamics of rare events have been studied in classical systems~\cite{HanggiTalknerBorkovec1990RMP,DellagoBolhuisGeissler2002ACP,BolhuisChandlerDellagoGeissler2002TPSReview}, and many of the techniques developed there have been  extended to Markovian open quantum systems. Large-deviation methods use a statistical bias, called a tilt, to give more weight to selected fluctuations and determine their probabilities~\cite{GarrahanLesanovsky2010LargeDeviations,CarolloGarrahanJack2021Level25,PerfettoCarolloLesanovsky2022Resetting}. Quantum cloning algorithms provide a way to sample these biased statistics by evolving a population of system copies and replicating or removing copies according to the bias~\cite{CarolloPerezEspigares2020QuantumCloning,LiuGu2025FirstPassageCloning}. A related approach, the quantum Doob transformation, constructs modified dynamics in which the selected rare fluctuations become typical~\cite{GarrahanLesanovsky2010LargeDeviations,CarolloEtAl2018QuantumDoob}. Transition path sampling (TPS) samples rare transitions through a Monte Carlo (MC) chain of trajectories without requiring the transition state or mechanism to be specified in advance~\cite{DellagoBolhuisCsajkaChandler1998TPS,DellagoBolhuisGeissler2002ACP}. It has been applied to quantum jump trajectories~\cite{SchileLimmer2018QuantumJumpTPS} and stochastic Schr\"odinger equations with Markovian noise~\cite{ChristieBolhuisLimmer2025QuantumSSE}, and the related transition path theory has been used to analyze reactive pathways of quantum jump trajectories~\cite{AndersonSchileLimmer2022QuantumJumpPaths}.

In many real-world examples, open quantum systems in which rare events occur are coupled to non-Markovian baths, for which the rare-event sampling methods above cannot be applied directly; examples include quantum Brownian motion~\cite{HanggiGrabertIngoldWeiss1985Memory,HanggiTalknerBorkovec1990RMP,HuPazZhang1992QBM,CaldeiraLeggett1983Tunneling}, Kerr-cat and superconducting qubits~\cite{FrattiniEtAl2024KerrCat,DubovitskiiEtAl2024Quasiparticle}, photon counting statistics~\cite{XuLi2018NonMarkovianCounting}, and work and heat fluctuations~\cite{DibaEtAl2024StrongCouplingWork,AurellDonvilMallick2020HeatLD}.

Here, we enable rare-event sampling outside the Markovian and perturbative regimes by formulating TPS for the formally exact non-Markovian quantum state diffusion (NMQSD) equation of motion (NMQSD-TPS)~\cite{DiosiStrunz1997NMSSE,DiosiGisinStrunz1998NMQSD,StrunzDiosiGisin1999NMQT}. In NMQSD, as in classical Langevin dynamics, each trajectory is a deterministic function of its initial state and a sequence of Gaussian random numbers. We therefore extend the noise-space TPS of Crooks and Chandler~\cite{CrooksChandler2001StochasticTPS} to allow for correlated noise arising from a non-Markovian bath. We demonstrate the formalism by calculating the tunneling rate for the normal-to-superradiant transitions of the bistable unbalanced open Dicke model, which become exponentially rare as the number of atoms $N$ in the cavity increases. At small $N$, these transitions are frequent enough to be observed directly in single quantum trajectories~\cite{StitelyEtAl2020Switching}, whereas their rates at large $N$ have previously been inferred by extrapolating brute-force results from smaller systems~\cite{MullerStrunz2025Dicke}. NMQSD-TPS determines these rates directly up to $N=500$.

\ssection{Transition path sampling of non-Markovian quantum trajectories}
We consider a system coupled to a non-Markovian Gaussian bosonic bath. In the interaction picture with respect to the bath Hamiltonian $\hat H_{\rm B}=\sum_\nu\omega_\nu\adag{\nu}\aop{\nu}$, the total Hamiltonian is given by ($\hbar=1$)
\begin{equation}
\hat H(t)=\hat H_{\rm sys}
+\hat L\hat B^\dagger(t)+\hat L^\dagger\hat B(t),
\label{eq:hamiltonian}
\end{equation}
where $\hat H_{\rm sys}$ is the system Hamiltonian and $\hat B(t)=\sum_\nu g_\nu e^{-i\omega_\nu t}\aop{\nu}$ is the bath operator. The bosonic operators $\aop{\nu}$ and $\adag{\nu}$ annihilate and create an excitation in mode $\nu$ with frequency $\omega_\nu$. The bath couples to the system through the system operator $\hat L$, with strength $g_\nu$ for mode $\nu$. The bath is initially in its vacuum state and uncorrelated with the system. The influence of this Gaussian bath on the system at zero temperature is determined by its vacuum correlation function,
\begin{equation}
\alpha(t)
=\expec{\hat B(t)\hat B^\dagger(0)}_{\rm vac}
=\sum_\nu |g_\nu|^2 e^{-i\omega_\nu t}.
\label{eq:bcf}
\end{equation}
We solve the NMQSD equation to obtain the reduced system dynamics for the Hamiltonian in \eref{eq:hamiltonian}~\cite{DiosiStrunz1997NMSSE,StrunzDiosiGisin1999NMQT},
\begin{equation}
\begin{aligned}
\partial_t\bm{\psi}(t,z^*)
={}&-i\hat H_{\rm sys}\bm{\psi}(t,z^*)
+\hat L z^*(t)\bm{\psi}(t,z^*)\\
&-\hat L^\dagger\int_0^t ds\,\alpha(t-s)
\frac{\delta\bm{\psi}(t,z^*)}{\delta z^*(s)},
\end{aligned}
\label{eq:nmqsd}
\end{equation}
where $\bm{\psi}(t,z^*)$ is a stochastic state of the system and $z(t)$ is a complex Gaussian noise with $\mathbb{E}[z(t)]=\mathbb{E}[z(t)z(s)]=0$ and $\mathbb{E}[z(t)z^*(s)]=\alpha(t-s)$. For a fixed initial state, $\bm{\psi}(t,z^*)$ depends on the noise values $z^*(s)$ for $0\le s\le t$, and the reduced density operator $\rho_{\rm sys}(t)={\rm Tr}_{\rm B}[\rho_{\rm tot}(t)]$ is recovered from the ensemble average over the noise $z(t)$,
\begin{equation}
\begin{aligned}
\rho_{\rm sys}(t)
&=\int\mathcal{D}z\,P[z]\,\ket{\bm{\psi}(t,z^*)}\bra{\bm{\psi}(t,z^*)}\\
&=\mathbb{E}[\ket{\bm{\psi}(t,z^*)}\bra{\bm{\psi}(t,z^*)}],
\end{aligned}
\label{eq:noise_average}
\end{equation}
where $P[z]$ is the Gaussian probability distribution of the noise. Here we have specialized these equations to a zero-temperature bath; the formalism carries over to thermal baths~\cite{SuessEisfeldStrunz2014HOPS,VarveloLyndBennett2021adHOPS,CittyEtAl2024MesoHOPS,LyndRaccah2025Peierls, DiosiGisinStrunz1998NMQSD}.

The hierarchy of pure states (HOPS) provides a numerical solution of the NMQSD equation~\cite{SuessEisfeldStrunz2014HOPS}. We use its nonlinear formulation, whose normalized stochastic states reproduce the reduced density operator through the same Gaussian average over the noise in \eref{eq:noise_average}~\cite{DiosiGisinStrunz1998NMQSD,VarveloLyndBennett2021adHOPS}. From this point onward, $\bm{\psi}(t,z^*)$ denotes these normalized states. For the present calculations, we use the nearly unitary hierarchy of pure states (nuHOPS), an extension of HOPS that improves the numerical efficiency for highly excited baths, such as the cavity mode in the Dicke model~\cite{MullerStrunz2025nuHOPS}. The nuHOPS equations and their relation to the standard HOPS hierarchy are given in the Supplemental Material, \sref{sup-sec:eom}. The code and data for the calculations in this work are available at Zenodo~\cite{PantZenodo2026}.

\begin{figure}[t]
\centering
\includegraphics[width=\figuresize]{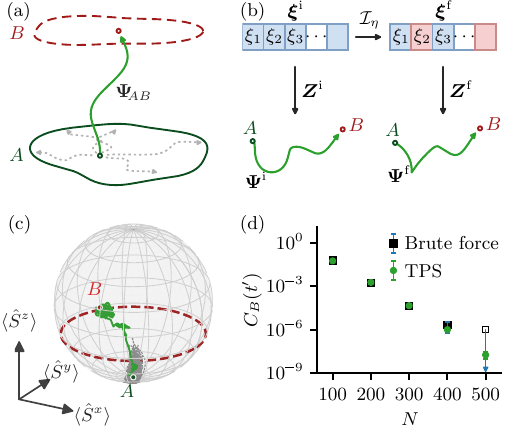}
\caption{NMQSD-TPS for normal-to-superradiant tunneling in the unbalanced open Dicke model. (a) A reactive trajectory $\bm{\Psi}_{AB}$ (green solid) connects the regions $A$ and $B$, while typical trajectories (gray dashed) return to $A$. (b) A TPS move replaces the independent Gaussian variables in a randomly chosen index set $\mathcal I_\eta$ (colored boxes), $\bm{\xi}^{\rm i}\to\bm{\xi}^{\rm f}$, which generates a new correlated noise sequence $\bm{Z}^{\rm i}\to \bm{Z}^{\rm f}$ and the corresponding trial trajectory $\bm{\Psi}^{\rm f}$. (c) Normal-to-superradiant tunneling on the collective-spin Bloch sphere. The dashed circle marks the mean-field superradiant value $\lam_{\rm SR}$, inside the region $B$ ($\lam\ge-0.6$). A reactive trajectory (green solid) connects the normal and superradiant regions, while most trajectories remain in the normal region (gray dashed lines). (d) The probability $C_B(t')$ that a trajectory is in $B$ at $t'=75\,\omega_a^{-1}$, from brute force (black squares) and TPS with WHAM (green circles), using the same number of trajectories for both methods at each $N$. Error bars show $\pm1$ bootstrap standard deviation. At $N=500$, none of the $10^6$ brute-force trajectories are in $B$; the open square with a downward arrow marks $10^{-6}$, one trajectory in $10^6$.
}
\label{fig:construction}
\end{figure}

We use TPS to sample rare transitions by constructing a Markov chain of reactive HOPS trajectories, which begin in an initial region $A$ of the system Hilbert space and end in a target region $B$, as illustrated in \frefp{fig:construction}{a}. To formulate TPS, we represent a HOPS trajectory by the sequence of system states at times $t_l=l\Delta t$, with $l=0,\ldots,L$,
\begin{equation}
\bm{\Psi}=(\bm{\psi}_0,\bm{\psi}_{t_1},\ldots,\bm{\psi}_{t_L}),
\qquad
\bm{\psi}_{t}=\bm{\psi}(t,z^*),
\label{eq:trajectory_def}
\end{equation}
and the corresponding sequence of correlated noise by
\begin{equation}
\bm Z=(z_0,z_{t_1},\ldots,z_{t_L}),
\qquad z_t=z(t).
\label{eq:noise_history_def}
\end{equation}
In the non-Markovian dynamics considered here, the values of $\bm Z$ are correlated in time. We therefore represent the noise by $\bm\xi$, the independent complex Gaussian variables (raw noise) from which $\bm Z$ is generated: a linear transformation obtained by circulant embedding maps $\bm\xi$ to $\bm Z$ with the required correlations~\cite{Percival2006Circulant,CittyEtAl2024MesoHOPS} (Supplemental Material, \sref{sup-sec:eom}). Therefore, for a fixed initial state ($\bm\psi_0$), $\bm\xi$ determines the noise sequence $\bm Z$ and the corresponding system trajectory $\bm\Psi$. 

At each MC step, a TPS move modifies the noise of the current trajectory to propose a new trajectory, called a trial trajectory~\cite{CrooksChandler2001StochasticTPS}. Because the correlated values of $\bm Z$ cannot be replaced independently, NMQSD-TPS moves act on $\bm\xi$, as in Monte Carlo sampling of rare fractional Brownian motion paths that avoid an absorbing boundary~\cite{HartmannMajumdarRosso2013fBm}. Starting from $\bm\xi^{\rm i}$ of the current reactive trajectory, we randomly select an index set $\mathcal I_\eta$ independent of the current noise values and replace the corresponding variables with new values from the same Gaussian distribution to obtain $\bm\xi^{\rm f}$, as illustrated in \frefp{fig:construction}{b}. The move size $\eta$ controls how many Gaussian variables are replaced (see \sref{sup-sec:seeds} in the Supplemental Material). Transforming $\bm\xi^{\rm f}$ gives the correlated noise $\bm Z^{\rm f}$ used to calculate the trial trajectory $\bm\Psi^{\rm f}$. The acceptance rule is chosen so that the reactive distribution of trajectories is the stationary distribution of the resulting Markov chain. For a fixed initial state $\bm\psi_0\in A$, the reactive distribution is (Supplemental Material, \eref{sup-eq:pi_ab})
\begin{equation}
\Pi_{AB} [\bm{\xi}] = \frac{1}{\mathcal Z_{AB}} \, h_A(\bm{\psi}_{0}) \, P[\bm{\xi}] h_B(\bm{\psi}_{t_L}),
\label{eq:pi_ab}
\end{equation}
where $h_A$ ($h_B$) is 1 if the trajectory starts (ends) in region $A$ ($B$) and zero otherwise, $P[\bm{\xi}]$ is the Gaussian probability distribution of $\bm{\xi}$, and the normalization constant $\mathcal Z_{AB}$ is the probability that a trajectory from the fixed initial state ends in $B$. Because the replaced variables are drawn from their own Gaussian distribution and $\mathcal I_\eta$ does not depend on $\bm\xi$, the noise probabilities of the current and trial trajectories cancel in the detailed-balance condition for \eref{eq:pi_ab}, which gives the acceptance rule (Supplemental Material, \eref{sup-eq:acceptance_final})
\begin{equation}
p_a(\bm{\xi}^{\rm i}\!\to\!\bm{\xi}^{\rm f})=h_B(\bm{\psi}_{t_L}^{\rm f}).
\label{eq:acceptance}
\end{equation}
A trial is therefore accepted if it ends in $B$; otherwise, the current trajectory is retained. After an initial equilibration, repeating these moves therefore samples the reactive ensemble.

\ssection{Dicke normal-to-superradiant tunneling}
The unbalanced open Dicke model describes $N$ identical two-level atoms, forming a collective spin $S=N/2$, coupled to a lossy cavity mode~\cite{MullerStrunz2025Dicke}. In the interaction picture of \eref{eq:hamiltonian}, the Hamiltonian is written as
\begin{equation}
\hat H(t)=\hat H_{\rm sys}
+\gbar\big[\hat L\hat a^\dagger(t)+\hat L^\dagger\hat a(t)\big],
\label{eq:dicke_hamiltonian_bath}
\end{equation}
with
\begin{equation}
\hat H_{\rm sys}=\omega_a\sz,
\qquad
\hat L=\frac{1}{\sqrt N}
\left(
\sx-i\frac{\dg}{\gbar}\sy
\right).
\label{eq:dicke_coupling}
\end{equation}
Here $\omega_a$ and $\omega_c$ are the atomic transition and cavity-mode frequencies, $\hat a(t)$ is the cavity annihilation operator in the interaction picture given in \eref{sup-eq:cavity_free_evolution} of the Supplemental Material, and $\hat S^\mu$ ($\mu=x,y,z$) are the collective spin components. The parameters $\gbar = g_{-} + g_{+}$ and $\dg = g_{-} - g_{+}$ are defined in terms of the rotating and counter-rotating couplings $g_{-}$ and $g_{+}$, respectively. The damped cavity mode acts as a non-Markovian bath for the collective spin, with vacuum correlation function $\alpha(t)=\gbar^2e^{-(\kappa+i\omega_c)t}$ for $t\ge0$, where $\kappa$ is the cavity amplitude decay rate~\cite{MullerStrunz2025nuHOPS, MullerStrunz2025Dicke}. We calculate the spin dynamics by solving the nuHOPS equations for this correlation function (see \sref{sup-sec:eom} of the Supplemental Material).

In the bistable regime, where the normal and superradiant states are both stable steady states of the mean-field dynamics, quantum fluctuations at finite $N$ can drive transitions between them. We use the parameters of Ref.~\cite{MullerStrunz2025Dicke}, $\omega_a = \omega_c = \kappa = 1$, $g_{-} = 1.8$, and $g_{+} = 0.782$, for which the normal-to-superradiant tunneling becomes exponentially rare as $N$ increases. We characterize this tunneling with the order parameter $\lam(t) = \expec{\sz}_t/S$, where $\expec{\hat O}_t=\bra{\bm{\psi}(t,z^*)}\hat O\ket{\bm{\psi}(t,z^*)}$ denotes the expectation value in a single trajectory. Every trajectory is initialized in the normal state $\ket{S, -S}$, for which $\lam(0) = -1$, so the initial region $A$ here reduces to this state. An example of a normal-to-superradiant transition is shown on the Bloch sphere in \frefp{fig:construction}{c}, where, after the transition, the trajectory fluctuates near the mean-field value $\lam_{\rm SR} \approx -0.412$.

At large $N$, too few trajectories reach the superradiant region for brute-force calculations to determine its population or the normal-to-superradiant tunneling rate. For these localized trajectories, the probability $C_B(t')$ that a trajectory is in the superradiant region $B$ at time $t'$ is the superradiant population of $\rho_{\rm sys}(t')$~\cite{MullerStrunz2025Dicke}. We define $B$ by $\lam\ge-0.6$, which includes the typical fluctuations around $\lam_{\rm SR}$. Figure~\ref{fig:construction}(d) shows the brute-force estimate of $C_B(t')$, the fraction of trajectories in $B$, at $t'=75\,\omega_a^{-1}$. At $N=400$, only two of $10^6$ trajectories are in $B$, giving $C_B(t')=(2.0\pm1.4)\times10^{-6}$, and none at $N=500$.

\ssection{Probability distribution}
\begin{figure*}[t]
\centering
\includegraphics[width=\textwidth]{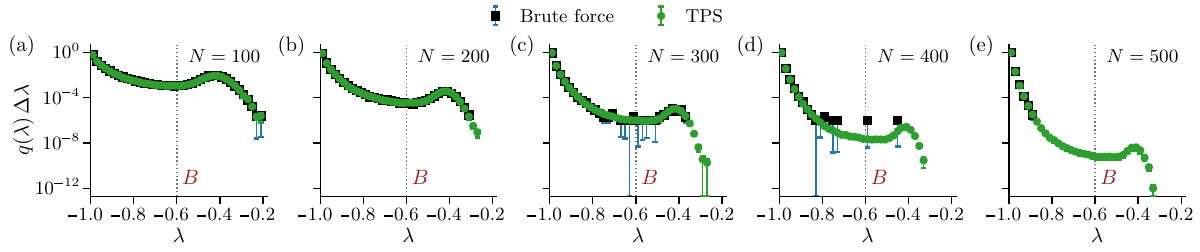}
\caption{TPS reconstructs the distribution of $\lam$ across the transition region, agreeing with brute force where brute force has enough trajectories and resolving the low-probability tail beyond the brute-force data. Probability per bin, $q_{t'}(\lambda)\Delta\lambda$, at $t'=75\,\omega_a^{-1}$ from TPS (green) and brute force (black) for (a) $N=100$, (b) $200$, (c) $300$, (d) $400$, and (e) $500$. The dotted line marks the boundary $\lambda_B=-0.6$ of the target region $B$. The bin width is $\Delta\lambda=0.02$. Error bars show $\pm1$ bootstrap standard deviation, obtained by resampling the independent TPS chains (green) and the brute-force trajectories (black). Both methods use the same number of trajectories at each $N$.}
\label{fig:window_wham}
\end{figure*}
TPS reconstructs the distribution $q_{t'}(\lam)$ of the order parameter at time $t'$ across the normal and superradiant regions. For a fixed initial state, the raw noise $\bm{\xi}$ determines the trajectory and hence the order parameter $\lam(t')$, so the distribution at time $t'$ is
\begin{equation}
q_{t'}(\lam)
=
\int \mathcal D\bm{\xi}\,
P[\bm{\xi}]\,
\delta\!\left[\lam-\lam(\bm{\psi}_{t'})\right].
\label{eq:fixed_time_distribution}
\end{equation}
To resolve the low-probability tail of this distribution, the range of $\lam(t')$ is divided into overlapping hard windows $W_i$~\cite{TorrieValleau1977Umbrella,DellagoBolhuisGeissler2002ACP}, and TPS samples trajectories for which $\lam(t')$ lies within the chosen window~\cite{DellagoBolhuisCsajkaChandler1998TPS}. The weighted histogram analysis method (WHAM)~\cite{KumarEtAl1992WHAM} combines the distributions obtained for these windows into the normalized $q_{t'}(\lam)$. The window acceptance rule, the WHAM equations and the sampling parameters are given in \sref{sup-sec:windows} and \sref{sup-sec:dicke_windows} of the Supplemental Material.

The reconstructed distribution from TPS agrees with brute-force calculations wherever brute force has enough trajectories, and it resolves the low-probability tail that brute force cannot. Figure~\ref{fig:window_wham} compares the two methods for $N=100$--$500$, using the same number of trajectories for each. For $N=100$--$300$, the distributions agree over their common range [\frefp{fig:window_wham}{a}--(c)]. For $N=400$, brute force gives only a few points beyond $\lam\approx-0.8$, with large relative uncertainties, and for $N=500$ none beyond $\lam\approx-0.88$, whereas TPS reconstructs the distribution across the transition and superradiant regions [\frefp{fig:window_wham}{d} and (e)].

TPS determines $C_B(t')$ beyond the reach of brute force, down to $10^{-8}$ at $N=500$. Integrating the reconstructed distribution over $B$ gives the values of $C_B(t')$ shown as green circles in \frefp{fig:construction}{d}. For $N=100$--$300$, the TPS values lie within 5\% of brute force. At $N=400$, TPS gives $C_B(t')=(1.0\pm0.3)\times10^{-6}$, consistent with the brute-force value $(2.0\pm1.4)\times10^{-6}$ but with a smaller uncertainty. At $N=500$, TPS gives $C_B(t')=(1.8\pm0.8)\times10^{-8}$, whereas none of the $10^6$ brute-force trajectories are reactive.

\ssection{Tunneling rate}
We obtain the tunneling rate from the slope of $C_B(t)=\int_B d\lam\,q_t(\lam)$, following the TPS rate method of Dellago, Bolhuis, and Chandler for classical dynamics~\cite{DellagoBolhuisChandler1999Rate,DellagoBolhuisGeissler2002ACP}. The method combines $C_B(t')$ at a single time, obtained by window sampling, with a second TPS ensemble that gives the time dependence of $C_B(t)$ without reconstructing $q_t(\lam)$ at every time. This ensemble consists of trajectories of duration $t_L>t'$ that start from the same initial state and visit $B$ at least once. The fraction $\mathbb{E}^{*}_{AB}\!\left[\hb(\bm{\psi}_{t})\right]$ of these trajectories in $B$ at time $t$ has the same time dependence as $C_B(t)$, and matching it to $C_B(t')$ at $t=t'$ fixes the constant factor between them (Supplemental Material, \sref{sup-sec:rate}). After an initial transient, $C_B(t)$ increases approximately linearly with $t$ [\fref{sup-fig:rate_plateau} of the Supplemental Material], and the rate is
\begin{equation}
\ktp=\frac{C_B(t')}{\mathbb{E}^{*}_{AB}\!\left[\hb(\bm{\psi}_{t'})\right]}\frac{d}{dt}\mathbb{E}^{*}_{AB}\!\left[\hb(\bm{\psi}_{t})\right]. \label{eq:rate_final}
\end{equation}
\begin{figure}[htb]
\centering
\includegraphics[width=\figuresize]{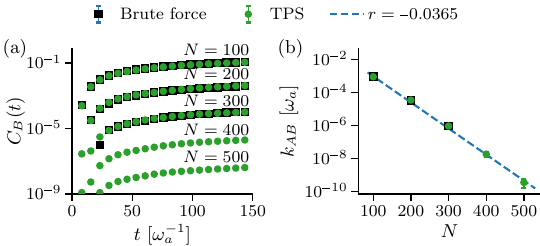}
\caption{TPS determines the tunneling rate up to $N=500$, where brute-force sampling no longer provides a reliable estimate. (a) Target population $C_B(t)$ from TPS (green) for $N=100$, $200$, $300$, $400$, and $500$ (top to bottom), with brute force (black) for $N\le300$. (b) Tunneling rate $\ktp$ as a function of $N$, from TPS (green) and brute force (black). The dashed line shows the exponential dependence $e^{r(N-100)}$ with $r=-0.0365$ from~\rref{MullerStrunz2025Dicke}, normalized to the brute-force rate at $N=100$. Error bars in (b) show $\pm1$ bootstrap standard deviation and are smaller than the symbols at $N\le300$.}
\label{fig:rate}
\end{figure}

TPS determines normal-to-superradiant tunneling rates up to $N=500$, beyond the range reliably accessible to brute force with the same number of trajectories. For $N\le300$, the reconstructed populations agree with brute force [\frefp{fig:rate}{a}], and the TPS rates lie within 10\% of the brute-force rates, inside their 10--25\% statistical uncertainties [\frefp{fig:rate}{b}]. Over $N=100$--$500$, the rates decrease approximately exponentially with $N$, with an exponent close to $r=-0.0365$, the slope between the rates at $N=50$ and $100$ in Fig.~3 of~\rref{MullerStrunz2025Dicke}. The dashed line in \frefp{fig:rate}{b} shows this exponential dependence, normalized to the brute-force rate at $N=100$.

\ssection{Conclusion}
In this work, we have formulated NMQSD-TPS for rare transitions in non-Markovian open quantum systems. The memory of the bath correlates the stochastic driving in time, so its values cannot be replaced independently, as in TPS for Markovian stochastic dynamics~\cite{CrooksChandler2001StochasticTPS}. We instead perform the TPS move in the independent complex Gaussian variables from which the NMQSD noise is generated~\cite{HartmannMajumdarRosso2013fBm}, so that the noise used to propagate the trial trajectory retains the correlations of the bath. The acceptance rule makes the reactive ensemble the stationary distribution of the Markov chain. 
Applied to normal-to-superradiant tunneling in the open Dicke model, NMQSD-TPS agrees with brute force within statistical uncertainties for $N\le300$ and determines superradiant populations down to $10^{-8}$ and tunneling rates up to $N=500$.

Unlike classical trajectories, the trajectories in the NMQSD-TPS ensemble are a scaffold for calculating averages over the noise and do not correspond to the record of an individual experiment. The HOPS state $\bm\psi(t,z^*)$ is the system state conditioned on a measurement of the bath at time $t$~\cite{DiosiStrunz1997NMSSE,GambettaWiseman2002RealNoise}, so the ensemble at each time describes a specific measurement. The state at a later time, however, assumes that the bath was not measured earlier, so the connection between times along a trajectory has no operational meaning~\cite{GambettaWiseman2002RealNoise,WisemanGambetta2008NoTrajectories}. TPS conditions on whole trajectories only as an intermediate step, and the quantities it determines refer to a single time, such as the probability that the measurement at time $t$ leaves the system in a given region. When the trajectories localize near metastable states, this probability equals the population of the corresponding state in $\rho_{\rm sys}(t)$, independent of the measurement, and its time dependence gives the transition rates between these states.

Finally, the NMQSD-TPS formalism presented here applies to any system Hamiltonian and coupling operator for which the reduced dynamics can be calculated with nonlinear HOPS. This includes systems with several sites and baths, as well as thermal baths with a Hermitian coupling operator, $\hat L=\hat L^\dagger$, beyond the vacuum bath considered in this work~\cite{SuessEisfeldStrunz2014HOPS,VarveloLyndBennett2021adHOPS,CittyEtAl2024MesoHOPS,LyndRaccah2025Peierls}. Rare barrier crossing in quantum Brownian motion and rare trajectories of exciton transfer, so far studied with Markovian treatments~\cite{ChristieBolhuisLimmer2025QuantumSSE,HosseinNejadEtAl2013ExcitonTransport}, are natural applications in which the temporal correlations of the bath can be retained. The same sampling principle could also be extended to rare photon counts and first-passage times~\cite{XuLi2018NonMarkovianCounting,MenczelEtAl2026FirstPassage}, and to work and heat fluctuations in non-Markovian quantum thermodynamics~\cite{FunoQuan2018PathIntegral,DibaEtAl2024StrongCouplingWork,GongEtAl2022WorkDistribution,PopovicEtAl2021HeatFCS,AurellDonvilMallick2020HeatLD}. Whether TPS outperforms brute force in these applications remains open and depends on Monte Carlo moves in the noise that decorrelate successive trajectories efficiently~\cite{GingrichGeissler2015NoiseGuidance}.

\begin{acknowledgments}
\ssection{Acknowledgments}
The authors acknowledge support from the Robert A. Welch Foundation (Grant No. F-2207-20240404). R.P. acknowledges support from the Texas Quantum Institute (TQI) Postdoctoral Fellowship at The University of Texas at Austin. We used Claude and ChatGPT to help develop the simulation code and plotting scripts. The authors verified the code and results and take full responsibility for this work.
\end{acknowledgments}

\bibliographystyle{apsrev4-2}
\makeatletter\immediate\write\@auxout{\string\citation{REVTEX42Control}}\immediate\write\@auxout{\string\citation{apsrev42Control}}\makeatother
\bibliography{references}

\end{document}


\title{Supplemental Material for:\\ Transition Path Sampling for Rare Events in Non-Markovian Open Quantum Systems}

\author{Ritesh Pant}
\affiliation{Department of Chemistry, University of Texas at Austin, Austin, TX 78712, USA}

\author{Kai Müller}
\affiliation{Institut für Theoretische Physik, Technische Universität Dresden, D-01062 Dresden, Germany}
\affiliation{Clarendon Laboratory, University of Oxford, Parks Road, Oxford OX1 3PU, United Kingdom}

\author{Tarun Gera}
\affiliation{Department of Chemistry, University of Texas at Austin, Austin, TX 78712, USA}

\author{Doran I.\ G.\ B.\ Raccah}
\affiliation{Department of Chemistry, University of Texas at Austin, Austin, TX 78712, USA}

\begin{abstract}
This Supplemental Material presents the equations used to calculate non-Markovian quantum trajectories and derives the acceptance rule for transition path sampling. We describe how to find a seed trajectory, reconstruct probability distributions from overlapping windows, and calculate rate constants. For the unbalanced open Dicke model, we also provide the sampling parameters, determine the burn-in of the Monte Carlo chains, and explain how the error bars and the comparisons with brute-force sampling are obtained.
\end{abstract}

\maketitle

\section{nuHOPS trajectory dynamics and noise construction}\label{sec:eom}

The memory term in the NMQSD formalism (\eref{main-eq:nmqsd} of the Letter) contains a functional derivative of the state with respect to the noise, so the equation cannot be propagated directly. HOPS handles this term by introducing auxiliary states $\bm{\Phi}^{(k)}(t,z^*)$ that obey a closed hierarchy of differential equations, and we use its near-unitary form, nuHOPS, to propagate the nonlinear, norm-preserving version of the NMQSD equation~\cite{SuessEisfeldStrunz2014HOPS,MullerStrunz2025nuHOPS,VarveloLyndBennett2021adHOPS}. For this, the bath correlation function is written as a sum of exponentials, and a single exponential suffices for the baths considered in this work, so that for $\tau\ge0$
\begin{equation}
\alpha(\tau)=G e^{-w\tau},
\qquad \alpha(-\tau)=\alpha^*(\tau),
\label{eq:alpha_modes}
\end{equation}
where $G$ is the weight of the exponential and $w$ is a complex parameter whose real and imaginary parts are the decay rate and the oscillation frequency of the bath correlation function. nuHOPS introduces a time-dependent shift $\mu(t)$ that accounts for the mean-field response of the bath, while the higher-order auxiliary states describe bath effects beyond the mean-field response. The shift is chosen such that the hierarchy stays near its zeroth-order state even for a highly excited bath~\cite{MullerStrunz2025nuHOPS}. The auxiliary states obey the nuHOPS equation
\begin{equation}
\begin{aligned}
\partial_t\bm{\Phi}^{(k)}(t,z^*)
={}&\Big[-i\hat H_{\rm sys}+\tilde z^*(t)\hat L-\mu(t)\hat L^\dagger-kw -\Gamma(t)\Big]
\bm{\Phi}^{(k)}(t,z^*)\\
&-i\sqrt{G(k+1)}
\big(\hat L^\dagger-\expec{\hat L^\dagger}_t\big)
\bm{\Phi}^{(k+1)}(t,z^*)
-i\sqrt{Gk}
\big(\hat L-\expec{\hat L}_t\big)
\bm{\Phi}^{(k-1)}(t,z^*),
\end{aligned}
\label{eq:nuhops_eom}
\end{equation}
where $k=0,1,\ldots$ is the hierarchy index of the auxiliary state, $\Gamma(t)$ is the normalization term, and $\expec{\hat O}_t$ is the expectation value in the normalized zeroth-order state,
\begin{equation*}
\expec{\hat O}_t=
\frac{\bra{\bm{\Phi}^{(0)}(t,z^*)}\hat O\ket{\bm{\Phi}^{(0)}(t,z^*)}}
{\braket{\bm{\Phi}^{(0)}(t,z^*)}{\bm{\Phi}^{(0)}(t,z^*)}},
\end{equation*}
and $\tilde z^*(t)=z^*(t)+\mu^*(t)$ is the shifted noise of nonlinear HOPS~\cite{SuessEisfeldStrunz2014HOPS,MullerStrunz2025nuHOPS}, with 
\begin{equation}
    \mu(t)=\int_0^t\alpha(t-s)\expec{\hat L}_s\,ds
\end{equation}
with time-evolution given by 
\begin{equation}
    \dot\mu(t)=-w\mu(t)+G\expec{\hat L}_t.
\end{equation}
nuHOPS also uses $\mu(t)$ to shift the auxiliary states, which gives the term $-\mu(t)\hat L^\dagger$ and the subtraction of $\expec{\hat L}_t$ in the coupling to $\bm{\Phi}^{(k-1)}(t,z^*)$; without these two terms, \eref{eq:nuhops_eom} reduces to nonlinear HOPS, up to the normalization of the auxiliary states~\cite{SuessEisfeldStrunz2014HOPS}.
The normalization term $\Gamma(t)$ preserves the norm of $\bm{\Phi}^{(0)}(t,z^*)$ and is given by~\cite{VarveloLyndBennett2021adHOPS}
\begin{equation}
\Gamma(t)=\mathrm{Re}\!\left[z^*(t)\expec{\hat L}_t\right]
-\frac{\mathrm{Re}\big[i\sqrt{G}\,\bra{\bm{\Phi}^{(0)}(t,z^*)}(\hat L^\dagger-\expec{\hat L^\dagger}_t)\ket{\bm{\Phi}^{(1)}(t,z^*)}\big]}
{\braket{\bm{\Phi}^{(0)}(t,z^*)}{\bm{\Phi}^{(0)}(t,z^*)}}.
\label{eq:nuhops_gauge}
\end{equation}

We integrate \eref{eq:nuhops_eom} with a fourth-order Runge--Kutta scheme ($\Delta t =10^{-3}\,\omega_a^{-1}$; $5\times10^{-4}\,\omega_a^{-1}$ for $N=500$). The hierarchy depth and the collective-spin state basis are constructed adaptively during each trajectory: the depth grows when the relative amplitude of the deepest auxiliary state exceeds $10^{-10}$ and shrinks when it falls below $10^{-12}$, between 4 and 600, and spin states at the edges of the kept range are added or removed when their relative norm exceeds $10^{-6}$ or falls below $10^{-8}$. The system state is the zeroth-order state, $\bm{\psi}(t,z^*)=\bm{\Phi}^{(0)}(t,z^*)$, and its values at the times $t_l$ form the trajectory $\bm{\Psi}$ of \eref{main-eq:trajectory_def} of the Letter.

The correlated noise $z(t)$ that drives \eref{eq:nuhops_eom} is generated from independent Gaussian variables by a fixed linear map. The noise sequence $\bm Z$ of \eref{main-eq:noise_history_def} of the Letter, whose values satisfy $\mathbb{E}[z(t_l)z^*(t_{l'})]=\alpha(t_l-t_{l'})$, is generated from a raw-noise vector $\bm\xi=(\xi_1,\ldots,\xi_{N_\xi})$ of $N_\xi$ independent unit complex Gaussian variables. The matrix $C$ maps $\bm\xi$ to the correlated noise and is chosen to reproduce the bath correlation function,
\begin{equation}
z(t_l)=\sum_{r=1}^{N_\xi}C_{lr}\,\xi_r,
\qquad
(CC^\dagger)_{ll'}=\alpha(t_l-t_{l'}).
\label{eq:filter_si}
\end{equation}
We construct $C$ by circulant embedding, which extends the bath correlation function periodically on a time grid longer than the trajectory interval $[0,t_L]$ so that the correlated noise $z(t_l)$ can be generated with fast Fourier transforms~\cite{Percival2006Circulant,CittyEtAl2024MesoHOPS}. The raw-noise vector $\bm\xi$ is therefore defined on this extended time grid, and we divide it into two blocks. The part of $\bm\xi$ corresponding to $[0,t_L]$ is the window block $\bm\xi^\sigma$, and the rest of $\bm\xi$, required for the extension, is the embedding block $\bm\xi^\theta$, so that $\bm\xi=(\bm\xi^\sigma,\bm\xi^\theta)$. The correlated noise $z(t_l)$ on $[0,t_L]$ depends on both blocks through \eref{eq:filter_si}.

For the Dicke model, the damped cavity mode is the non-Markovian bath of the collective spin, and its vacuum correlation function is a single exponential of the form of \eref{eq:alpha_modes}. The combined spin--cavity dynamics obeys a Markovian Lindblad master equation, whereas the spin alone is driven by the temporally correlated cavity field~\cite{MullerStrunz2025Dicke}. The bath operator in \eref{main-eq:dicke_hamiltonian_bath} of the Letter is $\hat B(t)=\gbar\hat a(t)$, and the bath correlation function is evaluated for the cavity and its vacuum loss field without the spin--cavity interaction, which enters separately through $\hat L$ in \eref{eq:nuhops_eom}. Without the spin, the cavity operator evolves under damping and vacuum input noise~\cite{GardinerCollett1985InputOutput},
\begin{equation}
\hat a(t)=e^{-(\kappa+i\omega_c)t}\hat a(0)
+\sqrt{2\kappa}\int_0^t ds\,
e^{-(\kappa+i\omega_c)(t-s)}\hat\zeta(s),
\label{eq:cavity_free_evolution}
\end{equation}
where $\hat\zeta(t)$ is the vacuum input field, with $[\hat\zeta(t),\hat\zeta^\dagger(s)]=\delta(t-s)$ and $\expec{\hat\zeta(t)\hat\zeta^\dagger(s)}_{\rm vac}=\delta(t-s)$. Here $\kappa$ is the decay rate of the cavity amplitude, corresponding to a photon-number decay rate $2\kappa$ in the absence of the atoms. The input noise preserves the cavity commutator $[\hat a(t),\hat a^\dagger(t)]=1$. With the cavity and input field initially in their joint vacuum state, \eref{eq:cavity_free_evolution} gives, for $t\ge s\ge0$,
\begin{equation}
\alpha(t-s)=\gbar^2\expec{\hat a(t)\hat a^\dagger(s)}_{\rm vac}
=\gbar^2e^{-(\kappa+i\omega_c)(t-s)}.
\label{eq:cavity_vacuum_correlation}
\end{equation}
This correlation function has the form of \eref{eq:alpha_modes} with $G=\gbar^2$ and $w=\kappa+i\omega_c$, so the damped cavity mode is treated with the single-exponential hierarchy of \eref{eq:nuhops_eom}~\cite{MullerStrunz2025nuHOPS}. In the calculations, we generate $z(t)$ on $[0,1.5\,t_L]$ and use only the values on $[0,t_L]$ in the nuHOPS equations. The circulant embedding doubles this interval to a periodic grid of length $3\,t_L$, so the window block $\bm\xi^\sigma$ covers $[0,t_L]$ and the embedding block $\bm\xi^\theta$ the remaining $2\,t_L$.

\section{Distribution of reactive trajectories and the acceptance rule}
\label{sec:tps}
In the Letter, \eref{main-eq:pi_ab} gives the distribution of reactive trajectories in terms of the raw noise and \eref{main-eq:acceptance} the acceptance rule. Here we derive this rule from detailed balance, starting from the probability of a trajectory~\cite{DellagoBolhuisGeissler2002ACP}. For a given initial state $\bm{\psi}_0$, the raw noise $\bm{\xi}$ determines the correlated noise through \eref{eq:filter_si}, and the nuHOPS equations driven by this noise determine the trajectory,
\begin{equation}
\bm{\Psi}(\bm{\xi}) = \mathcal{F}\big(\bm{\psi}_{0};\bm{Z}(\bm{\xi})\big),
\label{eq:trajectory_xi}
\end{equation}
where $\mathcal F$ denotes the time evolution under \eref{eq:nuhops_eom}.

The probability of a trajectory is the joint probability of its initial state and its noise sequence, which together specify $\bm\Psi$ through \eref{eq:trajectory_xi}~\cite{CrooksChandler2001StochasticTPS}. Assuming that the initial state and the noise are statistically independent, this probability factorizes,
\begin{equation}
P[\bm{\psi}_0,\bm Z] = P[\bm{\psi}_0]P[\bm Z],
\label{eq:path_probability}
\end{equation}
where $P[\bm{\psi}_{0}]$ is the distribution of initial states and $P[\bm Z]$ the probability density of the noise sequence. TPS samples reactive trajectories according to this probability, so the acceptance rule requires the noise probability density of the current and the trial trajectory.

We write this probability in terms of the raw noise $\bm\xi$. Since its variables are independent, the joint probability density of $\bm\xi$ is a product,
\begin{equation}
P[\bm\xi]=\prod_{r=1}^{N_\xi}\rho(\xi_r),
\label{eq:raw_measure}
\end{equation}
where $\rho(\xi_r)=\pi^{-1}e^{-|\xi_r|^2}$, so that $\mathbb{E}[\xi_r]=0$, $\mathbb{E}[\xi_r\xi_{r'}]=0$, and $\mathbb{E}[\xi_r\xi_{r'}^*]=\delta_{rr'}$. Since $\bm Z$ is a function of $\bm\xi$, we write the probability density of a trajectory in terms of $P[\bm\xi]$ from here on.

The reactive path ensemble is obtained by restricting $P[\bm\xi]$ to trajectories that start in $A$ and end in $B$ at time $t_L$~\cite{DellagoBolhuisGeissler2002ACP,ChristieBolhuisLimmer2025QuantumSSE}. We impose these conditions through the indicator functions
\begin{equation}
h_A(\bm{\psi}) = \begin{cases}
1,&\bm{\psi}\in A,\\
0,&\bm{\psi}\notin A
\end{cases},
\qquad
h_B(\bm{\psi}) = \begin{cases}
1,&\bm{\psi}\in B,\\
0,&\bm{\psi}\notin B
\end{cases}.
\label{eq:ha_hb}
\end{equation}
For a fixed initial state $\bm{\psi}_0\in A$, the probability density of reactive trajectories is then
\begin{equation}
\Pi_{AB}[\bm{\xi}] = \frac{h_A(\bm{\psi}_{0}) \, P[\bm{\xi}] \, h_B(\bm{\psi}_{t_L})}{\mathcal Z_{AB}},
\label{eq:pi_ab}
\end{equation}
where the normalization factor
\begin{equation}
\mathcal Z_{AB} = \int \mathcal D\bm{\xi} \, h_A(\bm{\psi}_{0}) \, P[\bm{\xi}] \, h_B(\bm{\psi}_{t_L})
\end{equation}
is the probability that a trajectory starting from $\bm{\psi}_0$ is in $B$ at $t_L$, with $\mathcal D\bm\xi=\prod_{r=1}^{N_\xi}d\,{\rm Re}\,\xi_r\,d\,{\rm Im}\,\xi_r$. For a rare transition, brute-force simulation requires on average $1/\mathcal Z_{AB}$ independent trajectories to obtain one reactive trajectory. TPS instead starts from one reactive trajectory and generates further reactive trajectories through successive Monte Carlo steps, each of which proposes a trial trajectory by a TPS move and accepts or rejects it. The resulting sequence of trajectories is a chain.

\ssection{TPS move}
\label{sec:mc_move}
The TPS move used in this work proposes a trial trajectory by replacing selected independent Gaussian variables in the raw-noise vector $\bm\xi$, as illustrated in \frefp{main-fig:construction}{b} of the Letter. The correlated noise sequence is recalculated from the modified variables with \eref{eq:filter_si}, and the trial trajectory is obtained by solving \eref{eq:nuhops_eom} with this noise. Changing only part of the raw noise can keep the trial trajectory close to the current reactive trajectory, so that it is more likely to end in $B$ than if all raw-noise variables were replaced~\cite{GingrichGeissler2015NoiseGuidance}.

The replacement is specified by an index set chosen independently of the current noise values. Let $\bm\xi^{\rm i}$ and $\bm\xi^{\rm f}$ denote the raw-noise vectors of the current and trial trajectories, with variables $\xi_r^{\rm i}$ and $\xi_r^{\rm f}$. We select a set $\mathcal I_\eta$ of indices $r$ with probability $p_{\rm sel}(\mathcal I_\eta)$, redraw the selected variables from $\rho(\xi_r)$, and leave all variables outside $\mathcal I_\eta$ unchanged. The generation probability density of this move is
\begin{equation}
p_g(\bm\xi^{\rm i}\!\to\!\bm\xi^{\rm f})
=\sum_{\mathcal I_\eta}p_{\rm sel}(\mathcal I_\eta)
\prod_{r\in\mathcal I_\eta}\rho(\xi_r^{\rm f})
\prod_{r\notin\mathcal I_\eta}
\delta^{(2)}(\xi_r^{\rm f}-\xi_r^{\rm i}),
\label{eq:p_forward}
\end{equation}
where the sum runs over all possible index sets and the first product is the probability density of the newly drawn values. In the second product, $\delta^{(2)}(u)=\delta({\rm Re}\,u)\delta({\rm Im}\,u)$ requires both the real and imaginary parts of every unselected variable to remain unchanged. The reverse generation probability density follows by exchanging the current and the trial raw noise,
\begin{equation}
p_g(\bm\xi^{\rm f}\!\to\!\bm\xi^{\rm i})
=\sum_{\mathcal I_\eta}p_{\rm sel}(\mathcal I_\eta)
\prod_{r\in\mathcal I_\eta}\rho(\xi_r^{\rm i})
\prod_{r\notin\mathcal I_\eta}
\delta^{(2)}(\xi_r^{\rm i}-\xi_r^{\rm f}).
\label{eq:p_reverse}
\end{equation}
\ssection{Detailed balance and the acceptance rule}
\label{sec:detail_balance}
The acceptance rule must preserve the reactive distribution $\Pi_{AB}[\bm{\xi}]$ under the Monte Carlo steps. We impose this condition through detailed balance between the current and trial trajectories~\cite{DellagoBolhuisGeissler2002ACP},
\begin{equation}
\Pi_{AB}[\bm{\xi}^{\rm i}]\,p_g(\bm{\xi}^{\rm i}\!\to\!\bm{\xi}^{\rm f})p_a(\bm{\xi}^{\rm i}\!\to\!\bm{\xi}^{\rm f})
=\Pi_{AB}[\bm{\xi}^{\rm f}]\,p_g(\bm{\xi}^{\rm f}\!\to\!\bm{\xi}^{\rm i})p_a(\bm{\xi}^{\rm f}\!\to\!\bm{\xi}^{\rm i}),
\label{eq:detail_balance}
\end{equation}
where $p_a(\bm{\xi}^{\rm i}\!\to\!\bm{\xi}^{\rm f})$ is the probability to accept $\bm{\xi}^{\rm f}$ once it has been generated. Rearranging \eref{eq:detail_balance} gives a condition for the acceptance probability alone,
\begin{equation}
\frac{p_a(\bm{\xi}^{\rm i}\!\to\!\bm{\xi}^{\rm f})}{p_a(\bm{\xi}^{\rm f}\!\to\!\bm{\xi}^{\rm i})}=R,
\qquad
R=
\frac{\Pi_{AB}[\bm{\xi}^{\rm f}]\,
p_g(\bm{\xi}^{\rm f}\!\to\!\bm{\xi}^{\rm i})}
{\Pi_{AB}[\bm{\xi}^{\rm i}]\,
p_g(\bm{\xi}^{\rm i}\!\to\!\bm{\xi}^{\rm f})},
\label{eq:acceptance_condition}
\end{equation}
which fixes only the ratio of the forward and reverse acceptance probabilities. Accepting or rejecting a trial requires each of them separately. We take the largest pair allowed by \eref{eq:acceptance_condition}, which is the Metropolis--Hastings rule,
\begin{equation}
p_a(\bm{\xi}^{\rm i}\!\to\!\bm{\xi}^{\rm f})
=\min\{1,R\}.
\label{eq:metropolis_ratio}
\end{equation}
For $R<1$, the forward acceptance probability is $R$ and the reverse one is $1$. For $R\geq1$, they are $1$ and $1/R$, respectively. Larger values with the same quotient would exceed $1$, so this rule rejects as few trials as detailed balance permits~\cite{DellagoBolhuisGeissler2002ACP}.

For the TPS move, the Gaussian probability factors cancel from the acceptance ratio, leaving only the conditions on the trajectory. To see this, note that an index set contributes to \esref{eq:p_forward}{eq:p_reverse} only if it contains every index at which $\bm\xi^{\rm i}$ and $\bm\xi^{\rm f}$ differ, and then it contributes to both sums with the same $p_{\rm sel}(\mathcal I_\eta)$. For each such set, the ratio of the reverse to the forward term is $\prod_{r\in\mathcal I_\eta}\rho(\xi_r^{\rm i})/\rho(\xi_r^{\rm f})$, the same for every contributing set, so the ratio of the sums equals the ratio for any one of them. Substituting \eref{eq:pi_ab} with \eref{eq:raw_measure} and this ratio gives
\begin{equation}
R=
\frac{\ha(\bm\psi_0^{\rm f})\hb(\bm{\psi}_{t_L}^{\rm f})}
{\ha(\bm\psi_0^{\rm i})\hb(\bm{\psi}_{t_L}^{\rm i})}
\frac{\prod_r\rho(\xi_r^{\rm f})}{\prod_r\rho(\xi_r^{\rm i})}
\frac{\prod_{r\in\mathcal I_\eta}\rho(\xi_r^{\rm i})}
{\prod_{r\in\mathcal I_\eta}\rho(\xi_r^{\rm f})}.
\label{eq:expanded_ratio}
\end{equation}
The Gaussian factors for the replaced variables cancel between the raw-noise and generation probabilities, while the factors for the unselected variables are identical in the current and trial trajectories. Only the indicator functions remain, giving
\begin{equation}
R=
\frac{\ha(\bm{\psi}_{0}^{\rm f})\hb(\bm{\psi}_{t_L}^{\rm f})}
{\ha(\bm{\psi}_{0}^{\rm i})\hb(\bm{\psi}_{t_L}^{\rm i})},
\label{eq:indicator_ratio}
\end{equation}
where $h_A(\bm{\psi}_{0}^{\rm i})=h_A(\bm{\psi}_{0}^{\rm f})=1$ because every trajectory is propagated from the same state $\bm{\psi}_0$ in $A$, and $h_B(\bm{\psi}_{t_L}^{\rm i})=1$ since the current trajectory is reactive. Hence $R=h_B(\bm{\psi}_{t_L}^{\rm f})$, which is $0$ or $1$, so that $\min\{1,R\}=R$ and \eref{eq:metropolis_ratio} gives
\begin{equation}
p_a(\bm{\xi}^{\rm i}\!\to\!\bm{\xi}^{\rm f})
=\hb(\bm{\psi}_{t_L}^{\rm f}),
\label{eq:acceptance_final}
\end{equation}
which is \eref{main-eq:acceptance} of the Letter.

\section{TPS move, seed search, and move size}
\label{sec:seeds}
The TPS move of the Letter, \frefp{main-fig:construction}{b}, is specified by the number of raw-noise variables replaced in each of the two blocks of $\bm\xi$ defined in \sref{sec:eom}. In each Monte Carlo step, we randomly select an index set $\mathcal I_\eta$ containing $n_\sigma$ distinct indices from the window block $\bm\xi^\sigma$ and $n_\theta$ from the embedding block $\bm\xi^\theta$. Every index within a block is equally likely to be selected, and each selected variable is replaced by an independent sample from $\rho(\xi_r)$. The selection probability $p_{\rm sel}(\mathcal I_\eta)$ is therefore the same for the forward and reverse moves, as used in \eref{eq:expanded_ratio}.

We express these move sizes as the time spanned by the replaced variables, $\eta_\sigma=n_\sigma\Delta t_\xi$ and $\eta_\theta=n_\theta\Delta t_\xi$, written together as $\eta=(\eta_\sigma,\eta_\theta)$, where $\Delta t_\xi$ is the spacing of the time grid on which the correlated noise is generated. The selected variables are spread over each block rather than forming a contiguous interval, so a move replaces the fraction $\eta_\sigma/t_L$ of the window block and $\eta_\theta/(2t_L)$ of the embedding block. The trial trajectory generated by this move is accepted or rejected with the rule of \eref{eq:acceptance_final}, or, for the Dicke calculations, with its window and visit forms, \esref{eq:window_acceptance_si}{eq:visit_acceptance}. An accepted trial replaces the current trajectory in the chain, whereas a rejected trial leaves the current trajectory unchanged.

\ssection{Greedy seed search}
The initial reactive trajectory of a chain, called the seed, is found by a greedy search that brings a trajectory closer to the target region $B$ in successive trials~\cite{CrooksChandler2001StochasticTPS}. We measure this progress with a distance $d[\bm{\Psi}]$ defined through the order parameter $\lam$, with $B=\{\lam\ge\lam_B\}$. For a seed that must end in $B$, we use
\begin{equation}
d[\bm{\Psi}]
=\max\{\lambda_B-\lambda(\bm{\psi}_{t_L}),\,0\}
+\max\Big\{\lambda_B-\max_{0\le l\le L}\lambda(\bm{\psi}_{t_l}),\,0\Big\},
\label{eq:seed_distance}
\end{equation}
where $\lam(\bm{\psi}_{t_l})$ is the order parameter at time $t_l$. The first term is the distance of the order parameter at the final time, $\lam(\bm{\psi}_{t_L})$, from $\lam_B$ and is zero if the trajectory ends in $B$. The second term is the smallest distance of the order parameter from $\lam_B$ along the trajectory and is zero if the trajectory reaches $B$ at any time. The total distance $d$ therefore vanishes only when the trajectory ends in $B$.

The greedy search reduces this distance by retaining only trial trajectories that are closer to the target. The first trial is a trajectory calculated from independently sampled $\xi_r$. In each later trial, $n_{\rm s}$ randomly selected $\xi_r$ in the window block $\bm{\xi}^\sigma$ are replaced by new samples from $\rho(\xi_r)$, and the trial trajectory is calculated from the modified $\bm{\xi}$. The seed-search move size is $\eta_{\rm s}=n_{\rm s}\Delta t_\xi$. The trial replaces the current trajectory only if its distance $d$ is smaller. The search stops when $d=0$, and the resulting trajectory is the seed from which the chain starts.

\begin{figure}[htb]
\centering
\includegraphics[width=0.49\textwidth]{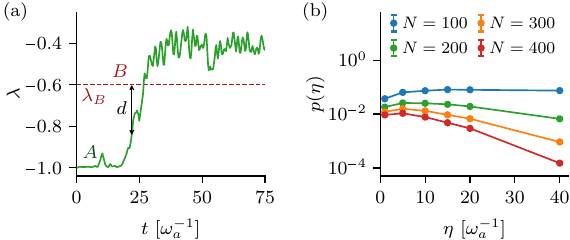}
\caption{Greedy seed search for the Dicke model. (a) Order parameter $\lam(t)$ of a reactive trajectory at $N=100$ that ends in $B$ at $t'=75\,\omega_a^{-1}$. The arrow schematically illustrates the distance of the order parameter from the target boundary $\lam_B$. (b) Fraction of trials that find a seed, \eref{eq:seed_probability}, from $576$ independent searches per point. Error bars show $\pm1$ bootstrap standard deviation, obtained by resampling the independent searches, and are smaller than the symbols.}
\label{fig:seed_eta}
\end{figure}

\ssection{Dicke model}
For the Dicke model of \eref{main-eq:dicke_hamiltonian_bath} of the Letter, with $N$ atoms and collective spin $S=N/2$, a reactive trajectory is one whose order parameter $\lam=\expec{\sz}_t/S$ starts in the normal state, $\lam(0)=-1$, and reaches the superradiant target region $B$, $\lam\ge\lam_B=-0.6$, as defined in the Letter. Figure~\ref{fig:seed_eta}(a) shows such a trajectory at $N=100$, which crosses $\lam_B$ and remains in $B$ at $t'=75\,\omega_a^{-1}$, so that it ends in $B$ and the distance of \eref{eq:seed_distance} vanishes. The arrow illustrates the distance of $\lam$ from $\lam_B$ in \eref{eq:seed_distance}.

We use the seed-search move size $\eta_{\rm s}=5\,\omega_a^{-1}$ in all our calculations because it gives the largest fraction of successful trials at $N=200$, $300$, and $400$, as shown in \frefp{fig:seed_eta}{b}. For each $N\in\{100,200,300,400\}$ and each $\eta_{\rm s}\in\{1,5,10,15,20,40\}\,\omega_a^{-1}$, we ran $N_{\rm s}=576$ independent greedy searches for a trajectory that ends in $B$ at $t'=75\,\omega_a^{-1}$, with a maximum of $5000$ moves per search, so that a search makes at most $N_{\rm trial}^{\max}=5001$ trials, counting the first. If $N_{\rm trial}$ is the total number of trials across these searches and $\bm{\psi}_{t'}^{(m)}$ is the state of trial $m$ at $t'$, the fraction of trials that find a seed is
\begin{equation}
p(\eta_{\rm s})=\frac{1}{N_{\rm trial}}
\sum_{m=1}^{N_{\rm trial}}\hb\!\left(\bm{\psi}_{t'}^{(m)}\right)
=\frac{n_{\rm found}}{N_{\rm trial}},
\label{eq:seed_probability}
\end{equation}
where $n_{\rm found}$ is the number of trials that end in $B$. Unsuccessful searches contribute all $5001$ trials to $N_{\rm trial}$ and no seed to $n_{\rm found}$, since excluding them would overestimate how frequently the search finds a seed. Such searches occurred only at $\eta_{\rm s}=40\,\omega_a^{-1}$, where $10$ of the $576$ searches at $N=300$ and $261$ of the $576$ searches at $N=400$ reached the trial limit without finding a seed.

\section{Hard window path sampling and WHAM method}
\label{sec:windows}
The reactive ensemble of \sref{sec:tps} gives the distribution of $\lam(t')$ for trajectories that end in $B$, but not the probability of the target region itself. We obtain that probability from the distribution of $\lam(t')$, \eref{main-eq:fixed_time_distribution} of the Letter, which is the normalized distribution over all trajectories started from $\bm{\psi}_0\in A$,
\begin{equation}
q_{t'}(\lambda)
=\int\mathcal D\bm{\xi}\,P[\bm{\xi}]\,
\delta\!\left[\lambda-\lambda(\bm{\psi}_{t'})\right],
\quad
\int\! d\lambda\,q_{t'}(\lambda)=1,
\label{eq:endpoint_density_si}
\end{equation}
where the trajectories are propagated only to $t'$, so that $t_L=t'$. The integral of $q_{t'}(\lam)$ over $B$ gives the probability of the target region at $t'$, $C_B(t')$ of the Letter. We therefore need $q_{t'}(\lam)$ over the whole range of $\lam(t')$ from $A$ to $B$, including near $B$ where it is small.

 We sample the whole range of $\lam(t')$ between $A$ and $B$ by dividing it into overlapping windows. For a rare transition, brute-force sampling gives $q_{t'}(\lam)$ accurately near $A$, but $q_{t'}(\lam)$ is hard to obtain near $B$, where it is many orders of magnitude smaller. We therefore choose overlapping windows $W_i=[w_i^-,w_i^+]$ covering the range of $\lam(t')$ from $A$ to $B$, and for each window, TPS samples trajectories whose $\lam(t')$ lies inside that window~\cite{TorrieValleau1977Umbrella,DellagoBolhuisCsajkaChandler1998TPS,DellagoBolhuisGeissler2002ACP}. The windows $W_i$ are intervals of the order parameter and are distinct from the window block $\bm{\xi}^\sigma$ of the raw noise in \sref{sec:eom}. The restriction to $W_i$ is expressed by
\begin{equation}
h_i(\lambda)=
\begin{cases}
1,&\lambda\in W_i,\\
0,&\text{otherwise},
\end{cases}
\label{eq:windows_def}
\end{equation}
in place of $h_B$ in \eref{eq:pi_ab}, so that the trajectories sampled for window $i$ are distributed with the probability density
\begin{equation}
\Pi_i[\bm{\xi}]
=\frac{h_A(\bm{\psi}_{0}) P[\bm{\xi}] h_i(\lambda(\bm{\psi}_{t'}))}{\mathcal Z_i},
\label{eq:window_ensemble}
\end{equation}
where $\mathcal Z_i=\int\mathcal D\bm{\xi}\,h_A(\bm{\psi}_{0})P[\bm{\xi}]h_i(\lambda(\bm{\psi}_{t'}))$ is the probability that $\lam(t')$ lies in $W_i$.

The trajectories sampled in window $i$ give $q_{t'}(\lam)$ inside $W_i$ up to the factor $\mathcal Z_i$. The restriction in \eref{eq:window_ensemble} removes the trajectories with $\lam(t')$ outside $W_i$ and renormalizes the rest, so the distribution of $\lam(t')$ in window $i$ is the conditional distribution
\begin{equation}
P_i(\lambda)
=\int \mathcal D\bm{\xi}\,\Pi_i[\bm{\xi}]\,
\delta\!\left[\lambda-\lambda(\bm{\psi}_{t'})\right]
=\frac{h_i(\lambda)\,q_{t'}(\lambda)}{\mathcal Z_i},
\label{eq:endpoint_conditional_density}
\end{equation}
which is $q_{t'}(\lam)$ restricted to $W_i$ and normalized within the window.

Window sampling uses the TPS move of \sref{sec:seeds} and the acceptance rule of \sref{sec:detail_balance} with $h_i$ in place of $h_B$, giving the acceptance condition
\begin{equation}
p_a^{W_i}(\bm{\xi}^{\rm i}\!\to\!\bm{\xi}^{\rm f})
=h_i(\lam(\bm{\psi}^{\rm f}_{t'})).
\label{eq:window_acceptance_si}
\end{equation}
A trial trajectory in window $W_i$ is therefore accepted if its $\lam(t')$ lies in $W_i$, and otherwise the current trajectory is retained. We use separate chains for each window, and the seed of a chain must have $\lam(t')$ in that window. We find this seed using the greedy search of \sref{sec:seeds}, with $d$ defined as the distance of $\lam(t')$ from $W_i$, which is zero inside the window.

 We estimate $P_i(\lam)$ from a histogram of the trajectories sampled in window $i$. Dividing $W_i$ into bins of width $\Delta\lam$ centered at $\lam_j$ gives
\begin{equation}
P_i(\lambda_j)=\frac{n_{ij}}{K_i\Delta\lambda},
\label{eq:window_histogram_estimate}
\end{equation}
where $n_{ij}$ is the number of Monte Carlo steps, excluding burn-in (\sref{sec:burn}), at which the current trajectory has $\lam(t')$ in bin $j$, and $K_i=\sum_j n_{ij}$ is the total number of these steps in window $i$. If $\mathcal Z_i$ were known, \esref{eq:endpoint_conditional_density}{eq:window_histogram_estimate} would give $q_{t'}(\lam)$ inside the window,
\begin{equation}
q_i(\lambda_j)
=\frac{\mathcal Z_i n_{ij}}{K_i\Delta\lambda}
\qquad \text{for }\lambda_j\in W_i.
\label{eq:single_window_q_estimate}
\end{equation}

We determine the unknown factors $\mathcal Z_i$ with the weighted histogram analysis method (WHAM), which combines the conditional distributions $P_i(\lam)$ of all windows into $q_{t'}(\lam)$~\cite{KumarEtAl1992WHAM}. WHAM uses the overlap of neighboring windows to relate their factors $\mathcal Z_i$. For a bin $j$ in the overlapping region, both estimates of \eref{eq:single_window_q_estimate} must give the same $q_{t'}(\lam_j)$, which fixes the ratio of the two $\mathcal Z_i$, and the normalization of $q_{t'}(\lam)$ fixes the remaining overall factor. Writing $f_i=\mathcal Z_i^{-1}$, the WHAM equations for hard windows are
\begin{equation}
q_{t'}(\lam_j)
=\frac{\sum_i n_{ij}}
{\Delta\lam\sum_i h_i(\lam_j)K_i f_i},
\label{eq:wham_estimate}
\end{equation}
and
\begin{equation}
f_i^{-1}=\mathcal Z_i
=\Delta\lam\sum_jh_i(\lam_j)q_{t'}(\lam_j),
\label{eq:wham_normalization}
\end{equation}
 which are solved iteratively: starting from positive trial values of $f_i$, we calculate $q_{t'}(\lam_j)$ using \eref{eq:wham_estimate} and update $\mathcal Z_i$ and $f_i$ using \eref{eq:wham_normalization}. These steps are repeated until $q_{t'}(\lam_j)$ and the factors $\mathcal Z_i$ stop changing. We then normalize the resulting distribution over the range covered by the windows, assuming negligible probability outside that range.

\section{Window sampling for the Dicke model}
\label{sec:dicke_windows}
\begin{figure*}[htb]
\centering
\includegraphics[width=\textwidth]{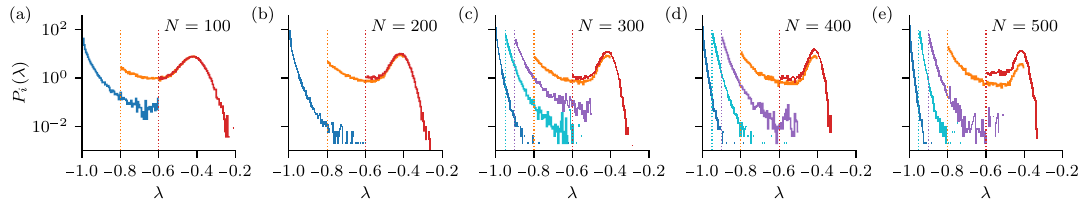}
\caption{Conditional distributions $P_i(\lam)$ of $\lam(t')$ in the windows $W_i$ for the Dicke model. (a) $N=100$, (b) $200$, (c) $300$, (d) $400$, and (e) $500$, with three windows at $N=100$ and $200$ and five at $N\ge300$. Dotted lines show the lower edge of each window. Windows $[-1.0,-0.6]$ (blue), $[-0.95,-0.55]$ (cyan), $[-0.9,-0.5]$ (purple), $[-0.8,-0.4]$ (orange), and $[-0.6,-0.2]$ (red).}
\label{fig:window_dists}
\end{figure*}

We next apply the window sampling and WHAM method to reconstruct the distribution $q_{t'}(\lam)$ at $t'=75\,\omega_a^{-1}$ for the Dicke model, shown in \fref{main-fig:window_wham} of the Letter. The windows are chosen to cover the range of $\lam(t')$ from $A$ ($\lam=-1$) to $B$ ($\lam\ge-0.6$), with sufficient overlap to allow WHAM to combine the distributions. The number and placement of windows depend on the system size $N$, as larger systems require more windows to capture the accurate distribution. At $N=100$ and $200$, we use three windows, $[-1.0,-0.6]$, $[-0.8,-0.4]$, and $[-0.6,-0.2]$. At $N\ge300$, we use five windows, $[-1.0,-0.6]$, $[-0.95,-0.55]$, $[-0.9,-0.5]$, $[-0.8,-0.4]$, and $[-0.6,-0.2]$. The conditional distributions $P_i(\lam)$ in each window for $N=100$--$500$ are shown in \fref{fig:window_dists}. The additional windows are needed at large $N$ because the distribution in $[-1.0,-0.6]$ becomes concentrated near $A$, leaving few samples in the interval shared with $[-0.8,-0.4]$. The window $[-0.6,-0.2]$ lies inside $B$, so every trajectory accepted in this window is reactive. The windows stop at $\lam=-0.2$ because $P_i(\lam)$ in $[-0.6,-0.2]$ falls to zero below this edge at every $N$ (\fref{fig:window_dists}), and none of the brute-force trajectories has $\lam(t')>-0.2$.

Each TPS move replaces about $13\%$ of the raw-noise variables in each block, with move sizes $(\eta_\sigma,\eta_\theta)=(10,20)\,\omega_a^{-1}$. The embedding block is twice as long as the window block, so $\eta_\theta=2\eta_\sigma$ replaces the same fraction in both blocks. Replacing fewer variables keeps the trial trajectory close to the current one, so $\lam(t')$ is more likely to remain in the window. Replacing more variables changes the trajectory more in each move, but also makes rejection more likely because $\lam(t')$ can leave the window. We chose these move sizes after comparing $\eta_\sigma\in\{5,10,20,40\}\,\omega_a^{-1}$, with $\eta_\theta=2\eta_\sigma$, in the window $[-0.6,-0.2]$ at $N=200$, with $96$ chains of $1500$ Monte Carlo steps for each move size, by the number of members per chain. Along a chain, we count a new member each time the accepted moves have replaced at least $90\%$ of the raw-noise variables since the previous member, so that successive members share at most $10\%$ of their raw noise. The number of members per chain therefore measures how quickly the accepted moves renew the raw noise. This number is largest for $(\eta_\sigma,\eta_\theta)=(10,20)\,\omega_a^{-1}$ and decreases for both smaller and larger move sizes.

\ssection{Burn-in}
\label{sec:burn}
The distribution of $\lam(t')$ obtained from the initial Monte Carlo steps can depend on the seed found by the greedy search~\cite{DellagoBolhuisGeissler2002ACP,CrooksChandler2001StochasticTPS}. WHAM combines the window distributions but does not remove this initial bias. We therefore determine a burn-in $m_{{\rm burn},i}$ for each window $W_i$ and discard the earlier Monte Carlo steps before calculating the window distributions.

Burn-in is needed only in the two upper windows, $[-0.8,-0.4]$ and $[-0.6,-0.2]$. We examine the dependence on the seed with $M=300$ independent chains in each of these windows. The chains used for WHAM in \tref{tab:window_sampling} are the first of these chains. Each chain starts from a separately obtained seed, and its Monte Carlo steps are grouped into ten consecutive blocks of equal length. At $N=100$, each chain has $1000$ Monte Carlo steps, giving $100$ steps per block. At $N=200$, each chain has $2500$ Monte Carlo steps, giving $250$ steps per block. At $N=300$, $400$, and $500$, each has $10000$ Monte Carlo steps, giving $1000$ steps per block.

The distribution in each block is calculated from the corresponding Monte Carlo steps across all $M$ independent chains. Let $m$ be the Monte Carlo step index in a chain and let $\mathcal B_\beta=[m_\beta,m_{\beta+1})$ denote the Monte Carlo steps in block $\beta$. For window $W_i$, we define
\begin{equation}
n_{ij}^{(\beta)}
=\sum_{c=1}^{M}\sum_{m\in \mathcal B_\beta}
b_j\!\left(\lam(\bm{\psi}^{(i,c,m)}_{t'})\right),
\label{eq:block_counts}
\end{equation}
where $\bm{\psi}^{(i,c,m)}_{t'}$ is the state at $t'$ of the trajectory at Monte Carlo step $m$ in independent chain $c$ of window $W_i$, and $b_j(\lam)=1$ if $\lam$ lies in bin $j$ and zero otherwise. Normalizing gives the block distribution,
\begin{equation}
p_{i,\beta}(j)=
\frac{n_{ij}^{(\beta)}}{\sum_{j'}n_{ij'}^{(\beta)}},
\label{eq:block_distribution}
\end{equation}
whose systematic change between successive blocks indicates dependence on the seed. Even when this dependence is no longer apparent, the distributions may still differ slightly because each block contains a finite number of samples.

\begin{figure*}[htb]
\centering
\includegraphics[width=\textwidth]{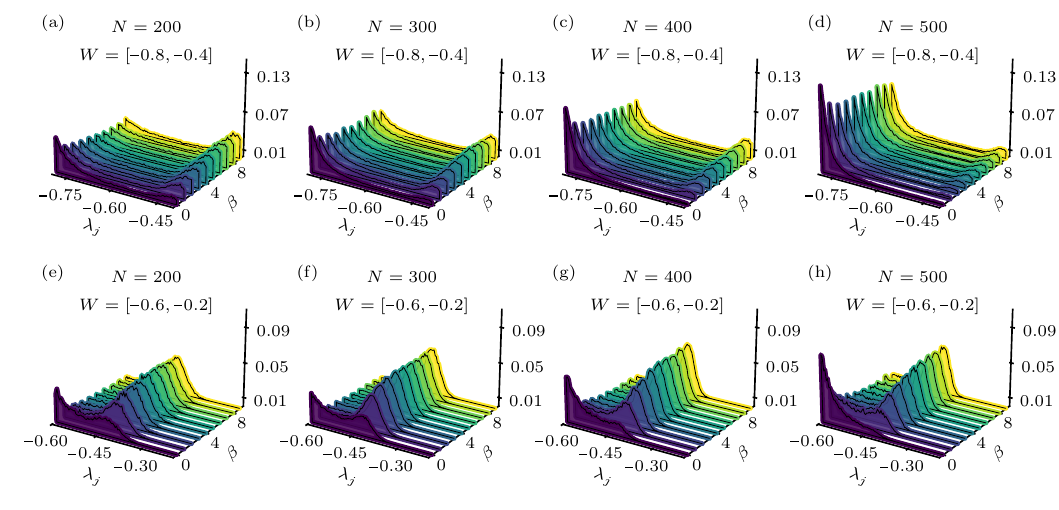}
\caption{Changes in the distribution of $\lam(t')$ over successive blocks of Monte Carlo steps in each window. The block distributions $p_{i,\beta}(j)$ are shown for $[-0.8,-0.4]$ (top row) and $[-0.6,-0.2]$ (bottom row) at $N=200$, $300$, $400$, and $500$ (columns), with the block index $\beta$ running from early to late along the depth. The Monte Carlo steps within each independent chain are grouped into ten consecutive blocks of equal length. Each curve shows the distribution of $\lam(t')$ for the same block of Monte Carlo steps across $300$ independent chains using bin width $\Delta\lam=0.004$.}
\label{fig:dist_mcblocks}
\end{figure*}

The distributions of the early blocks change in \fref{fig:dist_mcblocks}, while those of the late blocks approach the same shape. The figure shows the block distributions $p_{i,\beta}(j)$ of the two upper windows, with bins of width $\Delta\lam=0.004$. These changes motivate discarding the initial Monte Carlo steps. The distributions of the lower windows, $[-1.0,-0.6]$, $[-0.95,-0.55]$, and $[-0.9,-0.5]$, are not shown because we found no dependence on the seed in these windows.

We quantify the difference between two block distributions with the Hellinger distance defined as
\begin{equation}
H(p,p')=\sqrt{1-\sum_j\sqrt{p(j)p'(j)}},
\qquad 0\le H(p,p')\le1,
\label{eq:hellinger}
\end{equation}
which vanishes for identical distributions and becomes one when they do not overlap. The distributions of the last five blocks show no systematic change in \fref{fig:dist_mcblocks}, so we use their mean as the reference distribution $p_{i,\beta}^{\rm ref}(j)$ for the comparison. When examining a late block, we exclude that block from the mean to avoid making the distance artificially small. Therefore,
\begin{equation}
p_{i,\beta}^{\rm ref}(j)
=\frac{1}{N_{\rm ref}(\beta)}
\sum_{\substack{\beta'=6\\\beta'\ne\beta}}^{10}p_{i,\beta'}(j),\quad
d_i(\mathcal B_\beta)=H\big(p_{i,\beta},p_{i,\beta}^{\rm ref}\big),
\label{eq:block_reference}
\end{equation}
where $N_{\rm ref}(\beta)=5$ for $\beta=1,\ldots,5$ and $N_{\rm ref}(\beta)=4$ for $\beta=6,\ldots,10$. Even when the seed no longer affects the distribution, $d_i(\mathcal B_\beta)$ generally remains nonzero because each block contains a finite number of samples. Assuming that the late-block distributions are no longer affected by the seed, we use the last-block distance to estimate the differences caused by finite sampling.

\begin{figure*}[htb]
\centering
\includegraphics[width=\textwidth]{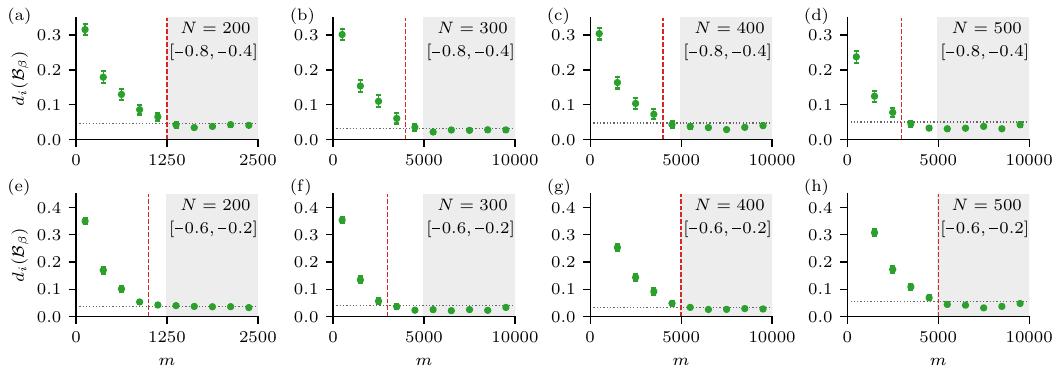}
\caption{Determining the burn-in from changes in the distribution of $\lam(t')$ over successive blocks of Monte Carlo steps. The Hellinger distance $d_i(\mathcal B_\beta)$ from the late-block reference is shown for $[-0.8,-0.4]$ (top row) and $[-0.6,-0.2]$ (bottom row) at $N=200$, $300$, $400$, and $500$ (columns). Error bars show $\pm1$ bootstrap standard deviation, obtained by resampling whole independent chains. The shaded region shows the late blocks used for the reference. The horizontal dotted line marks the upper end of the last block's error bar. The vertical dashed line indicates the burn-in step $m_{{\rm burn},i}$. The first block at $N=400$ and $500$ in the bottom row lies above the plotted range.}
\label{fig:hellinger_distance}
\end{figure*}

The Hellinger distances are shown in \fref{fig:hellinger_distance}. To estimate the uncertainty of these distances, we select $M$ whole chains with replacement from the $M$ independent chains. Each selected chain retains its full sequence of Monte Carlo steps, so correlations between successive trajectories are preserved. For each selection, we recalculate the block distributions, their references, and the distances. Repeating this procedure $4000$ times gives the variation in each distance, whose standard deviation is the error bar.

The burn-in is the start of the first block from which all later blocks agree with the last block within their error bars. That is, we choose the earliest block $\beta$ for which the lower end of every error bar from block $\beta$ onward lies at or below the upper end of the last block's error bar, as shown in \fref{fig:hellinger_distance} by the vertical red dashed line. The burn-in $m_{{\rm burn},i}$ is the Monte Carlo step at the start of this block, and the trajectories with $m<m_{{\rm burn},i}$ are discarded before applying WHAM. The burn-in for each $N$ is listed in \tref{tab:burnin}. At $N=100$, which is not shown in \fref{fig:hellinger_distance}, only the first block of $100$ Monte Carlo steps differs from the late blocks, and we discard it in both upper windows.

\begin{table}[htb]
\caption{Burn-in $m_{{\rm burn},i}$, in Monte Carlo steps, of the two upper windows. No Monte Carlo steps are discarded in the lower windows.}
\label{tab:burnin}
\centering
\begin{minipage}{0.5\textwidth}
\begin{ruledtabular}
\begin{tabular}{ccc}
$N$ & $[-0.8,-0.4]$ & $[-0.6,-0.2]$ \\
\hline
$100$ & $100$ & $100$ \\
$200$ & $1250$ & $1000$ \\
$300$ & $4000$ & $3000$ \\
$400$ & $4000$ & $5000$ \\
$500$ & $3000$ & $5000$ \\
\end{tabular}
\end{ruledtabular}
\end{minipage}
\end{table}

\ssection{Comparison with brute-force sampling}
The comparison in \fref{main-fig:window_wham} of the Letter uses the same number of trajectories for brute force and TPS, before discarding burn-in. Both methods use $4\times10^5$ trajectories at $N=100$ and $10^6$ at $N\ge200$, with the TPS trajectories divided among the windows as listed in \tref{tab:window_sampling}. The smaller number at $N=100$ suffices because brute force already gives many reactive trajectories there. Each TPS move generates one trial trajectory, so the numbers include rejected trials and trials during burn-in. The trajectories generated during the greedy seed searches of \sref{sec:seeds} are not included in these numbers. Including them, TPS uses at most $2\%$ more trajectories than brute force for every $N$, a negligible difference for this comparison. The two upper windows need more trajectories because their initial Monte Carlo steps are discarded as burn-in. The burn-in in \tref{tab:burnin} was determined from the $M=300$ independent chains per upper window of \sref{sec:burn}. The reconstruction uses the first chains of these, and the trajectories of the remaining chains are not counted in these numbers.

\begin{table}[htb]
\caption{Percentage of the total trajectories used for each window, Monte Carlo steps per chain, and number of independent chains. For each $N$, the percentages specify how the total number of trajectories is divided among the windows. These numbers include burn-in.}
\label{tab:window_sampling}
\centering
\small
\setlength{\tabcolsep}{3pt}
\begin{ruledtabular}
\begin{tabular}{ccccc}
$N$ & Window & \shortstack{Total\\trajectories\\(\%)} & \shortstack{Monte Carlo\\steps per chain} & \shortstack{Independent\\chains} \\
\hline
$100$ & $[-1.0,-0.6]$ & $10$ & $500$ & $80$ \\
 & $[-0.8,-0.4]$ & $45$ & $1000$ & $180$ \\
 & $[-0.6,-0.2]$ & $45$ & $1000$ & $180$ \\
\hline
$200$ & $[-1.0,-0.6]$ & $10$ & $1000$ & $100$ \\
 & $[-0.8,-0.4]$ & $45$ & $2500$ & $180$ \\
 & $[-0.6,-0.2]$ & $45$ & $2500$ & $180$ \\
\hline
$\ge300$ & $[-1.0,-0.6]$ & $10$ & $1000$ & $100$ \\
 & $[-0.95,-0.55]$ & $10$ & $1000$ & $100$ \\
 & $[-0.9,-0.5]$ & $10$ & $1000$ & $100$ \\
 & $[-0.8,-0.4]$ & $35$ & $10000$ & $35$ \\
 & $[-0.6,-0.2]$ & $35$ & $10000$ & $35$ \\
\end{tabular}
\end{ruledtabular}
\end{table}
 We reconstruct $q_{t'}(\lam)$ over $-1.0\le\lam\le-0.2$ with bins of width $\Delta\lam=0.005$ after discarding burn-in. For \fref{main-fig:window_wham} of the Letter, we combine four neighboring bins to show probabilities in bins of width $0.02$.

\ssection{Uncertainty of the reconstructed distribution}
The error bars in \fref{main-fig:window_wham} of the Letter estimate the statistical uncertainty in $q_{t'}(\lam)$ due to the finite number of sampled trajectories. Different chains start from separate greedy searches that use independent random numbers and are therefore independent, whereas the trajectories within a chain are correlated. We therefore estimate the uncertainty by resampling whole chains, as in the burn-in analysis, which preserves the correlations within each chain~\cite{HubDeGrootVanDerSpoel2010}.

For each window, we select whole chains with replacement from those used in the original reconstruction, using the number of chains listed in \tref{tab:window_sampling}. The selections are independent between windows. The Monte Carlo steps per chain and the discarded burn-in remain unchanged.

We calculate the conditional distribution $P_i(\lam)$ of each window from the selected chains with \eref{eq:window_histogram_estimate}, after discarding burn-in, and solve the WHAM equations again for $q_{t'}(\lam)$ and $\mathcal Z_i$. Repeating the chain selection and reconstruction gives a set of distributions whose variation determines the error bars. Each repetition is a bootstrap replicate~\cite{Efron1979}. If $q_{t'}^{(s)}(\lam_j)$ is the reconstructed density at $\lam_j$ in replicate $s$, the mean and variance over the replicates are
\begin{equation}
\bar q_{t'}(\lam_j)=\frac{1}{N_{\rm boot}}\sum_{s=1}^{N_{\rm boot}}q_{t'}^{(s)}(\lam_j),
\qquad
\sigma^2[q_{t'}(\lam_j)]
=\frac{1}{N_{\rm boot}-1}\sum_{s=1}^{N_{\rm boot}}
\left(q_{t'}^{(s)}(\lam_j)-\bar q_{t'}(\lam_j)\right)^2,
\label{eq:bootstrap}
\end{equation}
with $N_{\rm boot}=4000$ replicates. For \fref{main-fig:window_wham} of the Letter, we combine neighboring bins within each replicate and calculate the standard deviations of the resulting bin probabilities.

The same procedure gives the uncertainty in the probability of the target region at $t'$. We integrate $q_{t'}(\lam)$ over $B$ in each replicate and use the standard deviation of these probabilities for the TPS error bars in \frefp{main-fig:construction}{d} of the Letter.

For brute force, the trajectories are independent, so their uncertainty is estimated by resampling bin counts. We draw $4000$ new sets from the multinomial distribution defined by the observed bin probabilities, keeping the number of trajectories fixed. Their variation gives the brute-force error bars in \fref{main-fig:window_wham}, and the variation of the number of trajectories in $B$ gives those in \frefp{main-fig:construction}{d} of the Letter.

Using more chains gives the same distribution within the error bars. With $300$ independent chains in each upper window at each $N\ge200$ and the same number of Monte Carlo steps per chain as in \tref{tab:window_sampling}, the reconstructed distribution agrees with the original reconstruction within its error bars.

\section{Rate constant from transition path sampling}
\label{sec:rate}
Equation~(\ref{main-eq:rate_final}) of the Letter gives the tunneling rate from the probability of the target region at the fixed time $t'$ and from the time dependence of the fraction in $B$ of a second ensemble of trajectories, those that visit $B$ at least once~\cite{DellagoBolhuisChandler1999Rate,DellagoBolhuisGeissler2002ACP}. Here we derive this expression and explain how the rate shown in \fref{main-fig:rate} of the Letter is calculated for the Dicke model.

The rate follows from the time dependence of the probability of the target region. The probability that a trajectory initially in $A$ is in $B$ at time $t$ is
\begin{equation}
C_B(t)
=\frac{\mathbb{E}\!\left[\ha(\bm{\psi}_{0})\hb(\bm{\psi}_{t})\right]}
{\mathbb{E}\!\left[\ha(\bm{\psi}_{0})\right]},
\label{eq:cb_definition_si}
\end{equation}
where $\mathbb{E}[\cdot]$ denotes an average over the unbiased raw-noise distribution $P[\bm{\xi}]$.

The slope of $C_B(t)$ gives the forward tunneling rate after the initial transient and before depletion of $A$ or reverse transitions appreciably affect $C_B(t)$. We describe this regime by transitions between $A$ and $B$ with forward and reverse rate constants $k_{AB}$ and $k_{BA}$~\cite{BolhuisChandlerDellagoGeissler2002TPSReview,DellagoBolhuisCsajkaChandler1998TPS}. For trajectories initially in $A$, this approximation gives
\begin{equation}
C_B(t)\simeq \mathbb{E}_{\rm ss}\!\left[\hb\right]\big(1-e^{-t/\tau_{\rm rxn}}\big),
\qquad
\tau_{\rm rxn}^{-1}=k_{AB}+k_{BA},
\label{eq:two_state_relaxation}
\end{equation}
with $\mathbb{E}_{\rm ss}[h_B]=k_{AB}/(k_{AB}+k_{BA})$ the stationary population of $B$ in this approximation. At times short compared with $\tau_{\rm rxn}$, depletion of $A$ and reverse transitions have a small effect, and \eref{eq:two_state_relaxation} gives $dC_B/dt\simeq k_{AB}$. We therefore reconstruct $C_B(t)$ and identify an interval in which its slope is constant.

The WHAM reconstruction in \sref{sec:windows} gives the absolute probability at $t'$,
\begin{equation}
C_B(t')
=\int_B d\lambda\,q_{t'}(\lambda),
\label{eq:ct_prime_si}
\end{equation}
but this single value does not determine the time dependence of the population. Rather than reconstructing $q_t(\lam)$ separately at every time, we sample trajectories of duration $t_L>t'$ that enter $B$ at least once during $[0,t_L]$~\cite{DellagoBolhuisGeissler2002ACP,ChristieBolhuisLimmer2025QuantumSSE}. We express this visit condition by
\begin{equation}
H_B[\bm{\Psi}]
=\max_{0\le l\le L}h_B(\bm{\psi}_{t_l}),
\label{eq:visit_functional}
\end{equation}
which is one if any state along the trajectory lies in $B$, and zero otherwise. Restricting the unbiased trajectories to those satisfying this condition gives the visit ensemble,
\begin{equation}
\Pi_{AB}^{*}[\bm{\xi}]
=\frac{h_A(\bm{\psi}_{0}) P[\bm{\xi}] H_B[\bm{\Psi}]}
{\mathcal Z_{AB}^{*}},
\label{eq:visit_ensemble}
\end{equation}
where $\mathcal Z_{AB}^{*}=\int\mathcal D\bm{\xi}\,h_A(\bm{\psi}_{0})P[\bm{\xi}]H_B[\bm{\Psi}]$ is the probability of visiting $B$ within $t_L$ for our fixed initial state in $A$.

The visit ensemble uses the same TPS move as in \sref{sec:seeds}, with the acceptance condition changed to require a visit to $B$. Replacing $h_B(\bm{\psi}_{t_L})$ by $H_B[\bm\Psi]$ in the detailed-balance derivation of \sref{sec:detail_balance} gives
\begin{equation}
p_a^{\rm visit}(\bm{\xi}^{\rm i} \to \bm{\xi}^{\rm f}) = H_B[\bm{\Psi}^{\rm f}].
\label{eq:visit_acceptance}
\end{equation}
A trial trajectory is therefore accepted if it enters $B$ at least once; otherwise, the current trajectory is retained. The seed must satisfy the same visit condition. We therefore use only the distance $\max\{\lambda_B-\max_l\lambda(\bm{\psi}_{t_l}),0\}$ in the greedy search of \sref{sec:seeds}, corresponding to the second term of \eref{eq:seed_distance}.

The fraction of trajectories in $B$ is proportional to the unbiased probability $C_B(t)$. We calculate this fraction at each time $t$ as
\begin{equation}
\mathbb{E}^{*}_{AB}\!\left[h_B(\bm{\psi}_{t})\right]
=\frac{1}{\mathcal Z_{AB}^{*}}
\int\mathcal D\bm{\xi}\,P[\bm{\xi}]h_A(\bm{\psi}_{0})
H_B[\bm{\Psi}]h_B(\bm{\psi}_{t}),
\label{eq:visit_population_average}
\end{equation}
where $\mathbb{E}^{*}_{AB}[\cdot]$ denotes an average over trajectories that visit $B$ at least once during $[0,t_L]$. A trajectory that is in $B$ at time $t$ has necessarily visited $B$ during $[0,t_L]$, so at any sampled time $t\le t_L$,
\begin{equation}
h_B(\bm{\psi}_{t})H_B[\bm{\Psi}]=h_B(\bm{\psi}_{t}).
\label{eq:visit_identity}
\end{equation}
Substituting this identity into \eref{eq:visit_population_average} gives
\begin{equation}
\mathbb{E}^{*}_{AB}\!\left[h_B(\bm{\psi}_{t})\right]
=\frac{1}{\mathcal Z_{AB}^{*}}
\int\mathcal D\bm{\xi}\,P[\bm{\xi}]h_A(\bm{\psi}_{0})h_B(\bm{\psi}_{t})
=\frac{\mathbb{E}\!\left[h_A(\bm{\psi}_{0})\right]}{\mathcal Z_{AB}^{*}}\,C_B(t),
\label{eq:visit_to_unbiased}
\end{equation}
where the prefactor is the inverse probability that a trajectory starting in $A$ visits $B$ within $t_L$. This probability can be very small, but we do not need to calculate it. The prefactor is independent of $t$ and cancels when we divide \eref{eq:visit_to_unbiased} by the same equation evaluated at $t'$, which gives
\begin{equation}
C_B(t)
=C_B(t')
\frac{\mathbb{E}^{*}_{AB}\!\left[h_B(\bm{\psi}_{t})\right]}
{\mathbb{E}^{*}_{AB}\!\left[h_B(\bm{\psi}_{t'})\right]}.
\label{eq:rate_bridge}
\end{equation}
We identify an interval after the initial transient in which $C_B(t)$ increases linearly and $t\ll\tau_{\rm rxn}$, and the tunneling rate is the slope of $C_B(t)$ in this interval,
\begin{equation}
k_{AB}
=\frac{C_B(t')}
{\mathbb{E}^{*}_{AB}\!\left[h_B(\bm{\psi}_{t'})\right]}
\frac{d}{dt}\mathbb{E}^{*}_{AB}\!\left[h_B(\bm{\psi}_{t})\right],
\label{eq:rate_final_si}
\end{equation}
which is \eref{main-eq:rate_final} of the Letter.

\ssection{Dicke model}
For the Dicke model, we reconstruct $C_B(t)$ by combining the window sampling of \sref{sec:dicke_windows} with trajectories that visit $B$. The probability $C_B(t')$ is obtained at $t'=75\,\omega_a^{-1}$, and the time dependence comes from visit trajectories of duration $t_L=150\,\omega_a^{-1}$. Substituting both into \eref{eq:rate_bridge} gives $C_B(t)$. The normal-to-superradiant tunneling rate is the slope of $C_B(t)$ after the initial transient, where $C_B(t)$ is linear in $t$.

The visit trajectories are twice as long as the window-sampling trajectories, so we double both move sizes to keep the replaced fraction of raw noise unchanged. We use $\eta_\sigma=20\,\omega_a^{-1}$ and $\eta_\theta=40\,\omega_a^{-1}$, which replace about $13\%$ of the $\xi_r$ in each block in every move.

The rate comparison uses the same total number of trajectories for TPS and brute force, with the TPS trajectories divided between window and visit sampling as listed in \tref{tab:rate_sampling}. At $N=100$, window sampling and visit sampling use $2\times10^5$ trajectories each. At $N\ge200$, they use $8\times10^5$ and $2\times10^5$ trajectories. The seed searches for window and visit sampling are not counted in these numbers. Including them, TPS uses at most $4\%$ more trajectories than brute force for every $N$, which does not significantly affect the rate comparison. For every $N$, the visit ensemble uses $200$ independent chains of $1000$ Monte Carlo steps. Applied to $\lam(t')$ of these chains, the burn-in test of \sref{sec:burn} shows no early block that differs from the late blocks, so all Monte Carlo steps are used.

\begin{table}[htb]
\caption{Percentage of the total trajectories used for each window and for the visit ensemble in the rate comparison of \fref{main-fig:rate} of the Letter, Monte Carlo steps per chain, and number of independent chains. For each $N$, the percentages specify how the total number of trajectories, $4\times10^5$ at $N=100$ and $10^6$ at $N\ge200$, is divided. These numbers include burn-in.}
\label{tab:rate_sampling}
\centering
\small
\setlength{\tabcolsep}{3pt}
\begin{ruledtabular}
\begin{tabular}{ccccc}
$N$ & Window & \shortstack{Total\\trajectories\\(\%)} & \shortstack{Monte Carlo\\steps per chain} & \shortstack{Independent\\chains} \\
\hline
$100$ & $[-1.0,-0.6]$ & $10$ & $500$ & $80$ \\
 & $[-0.8,-0.4]$ & $20$ & $1000$ & $80$ \\
 & $[-0.6,-0.2]$ & $20$ & $1000$ & $80$ \\
 & visit ensemble & $50$ & $1000$ & $200$ \\
\hline
$200$ & $[-1.0,-0.6]$ & $10$ & $1000$ & $100$ \\
 & $[-0.8,-0.4]$ & $35$ & $2500$ & $140$ \\
 & $[-0.6,-0.2]$ & $35$ & $2500$ & $140$ \\
 & visit ensemble & $20$ & $1000$ & $200$ \\
\hline
$\ge300$ & $[-1.0,-0.6]$ & $10$ & $1000$ & $100$ \\
 & $[-0.95,-0.55]$ & $10$ & $1000$ & $100$ \\
 & $[-0.9,-0.5]$ & $10$ & $1000$ & $100$ \\
 & $[-0.8,-0.4]$ & $25$ & $10000$ & $25$ \\
 & $[-0.6,-0.2]$ & $25$ & $10000$ & $25$ \\
 & visit ensemble & $20$ & $1000$ & $200$ \\
\end{tabular}
\end{ruledtabular}
\end{table}

At $N\ge200$, window sampling uses more trajectories than visit sampling because the initial Monte Carlo steps in the upper windows are discarded as burn-in. The lower windows use the same chains as in \tref{tab:window_sampling}, and the remaining trajectories are divided equally between $[-0.8,-0.4]$ and $[-0.6,-0.2]$, with the burn-in values of \sref{sec:burn}. The rate uses the value of $C_B(t')$ obtained from these windows, which have fewer upper-window chains than the reconstruction in \fref{main-fig:window_wham} of the Letter: $C_B(t')=5.76\times10^{-2}$, $1.84\times10^{-3}$, $4.65\times10^{-5}$, $8.52\times10^{-7}$, and $1.86\times10^{-8}$ for $N=100$, $200$, $300$, $400$, and $500$. These values agree with those in \frefp{main-fig:construction}{d} of the Letter within their statistical uncertainties. As for \fref{main-fig:window_wham}, the upper-window chains are the first of the $M=300$ chains of \sref{sec:burn}, and the trajectories of the remaining chains are not counted.

\begin{figure*}[htb]
\centering
\includegraphics[width=\textwidth]{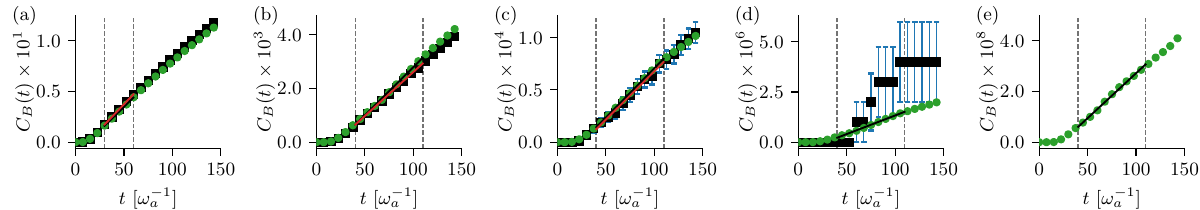}
\caption{Extracting the tunneling rate from $C_B(t)$. Green points show the TPS reconstruction from \eref{eq:rate_bridge} for (a) $N=100$, (b) $200$, (c) $300$, (d) $400$, and (e) $500$. The black line is the linear fit between the grey dashed lines. The fitted slope gives $k_{AB}$. Black squares show the brute-force $C_B(t)$, with $\pm1$ standard deviation error bars from resampling the brute-force trajectories. In (a)--(c), the red line is the brute-force fit over the same time interval. At $N=400$, only four brute-force trajectories are in $B$ at $t_L$, so no fit is shown. At $N=500$, none of the $10^6$ brute-force trajectories visits $B$ within $t_L$.}
\label{fig:rate_plateau}
\end{figure*}

The initial transient in $C_B(t)$ reflects the evolution from the normal region toward $B$. Every trajectory starts at $\lam=-1$ and contributes to the population of $B$ only when $\lam(t)\ge\lam_B$. The transient is longer at $N\ge200$ than at $N=100$, as seen in \fref{fig:rate_plateau}.

To identify the interval in which $C_B(t)$ increases linearly, we examine whether its slope remains constant in time. We found an approximately linear increase over $30\le\omega_a t\le60$ at $N=100$ and $40\le\omega_a t\le110$ at $N=200$--$500$. These intervals exclude the initial transient and the later departure from linear behavior. Over the same intervals, we fit the fraction of visit trajectories in $B$ by least squares to
\begin{equation}
\mathbb{E}^{*}_{AB}\!\left[h_B(\bm{\psi}_{t})\right]=a_0+a_1t,
\label{eq:sliding_linear_fit}
\end{equation}
and substitute the slope $a_1$ into \eref{eq:rate_final_si} to obtain the tunneling rate. The resulting rates are shown in \frefp{main-fig:rate}{b} of the Letter and listed with their uncertainties in \tref{tab:rates}. The uncertainties of the TPS rates are the standard deviations over $4000$ bootstrap replicates, each of which selects whole chains with replacement in every window, as for \eref{eq:bootstrap}, and independently among the $200$ visit chains, and repeats the WHAM reconstruction, the fit of \eref{eq:sliding_linear_fit}, and \eref{eq:rate_final_si}; for brute force, the trajectories are resampled and the fit of $C_B(t)$ is repeated.

\begin{table}[htb]
\caption{Tunneling rates $k_{AB}$ from TPS and from brute force, in units of $\omega_a$, with their statistical uncertainties ($\pm1$ standard deviation), as shown in \frefp{main-fig:rate}{b} of the Letter.}
\label{tab:rates}
\centering
\small
\begin{ruledtabular}
\begin{tabular}{ccc}
$N$ & $k_{AB}$ (TPS) & $k_{AB}$ (brute force) \\
\hline
\rule{0pt}{2.6ex}$100$ & $(9.32\pm0.89)\times10^{-4}$ & $(9.97\pm0.10)\times10^{-4}$ \\
$200$ & $(3.58\pm0.78)\times10^{-5}$ & $(3.27\pm0.08)\times10^{-5}$ \\
$300$ & $(9.40\pm2.39)\times10^{-7}$ & $(9.24\pm1.26)\times10^{-7}$ \\
$400$ & $(1.86\pm0.65)\times10^{-8}$ & --- \\
$500$ & $(3.53\pm1.87)\times10^{-10}$ & --- \\
\end{tabular}
\end{ruledtabular}
\end{table}

\bibliographystyle{apsrev4-2}
\makeatletter\immediate\write\@auxout{\string\citation{REVTEX42Control}}
\immediate\write\@auxout{\string\citation{apsrev42Control}}
\makeatother
\bibliography{references}